\documentclass[fleqn,usenatbib]{mnras}

\usepackage{newtxtext,newtxmath}

\usepackage[T1]{fontenc}

\DeclareRobustCommand{\VAN}[3]{#2}
\let\VANthebibliography\thebibliography
\def\thebibliography{\DeclareRobustCommand{\VAN}[3]{##3}\VANthebibliography}

\usepackage{graphicx}	
\usepackage{amsmath}	
\usepackage{bbding}	
\usepackage{color}
\usepackage{gensymb}
\usepackage{comment}
\usepackage{empheq}
\usepackage{comment}

\usepackage{scalerel,tikz}
\usetikzlibrary{svg.path}
\definecolor{orcidlogocol}{HTML}{A6CE39}
\tikzset{orcidlogo/.pic={
 \fill[orcidlogocol] svg{M256,128c0,70.7-57.3,128-128,128C57.3,256,0,198.7,0,128C0,57.3,57.3,0,128,0C198.7,0,256,57.3,256,128z};
 \fill[white] svg{M86.3,186.2H70.9V79.1h15.4v48.4V186.2z}
 svg{M108.9,79.1h41.6c39.6,0,57,28.3,57,53.6c0,27.5-21.5,53.6-56.8,53.6h-41.8V79.1z M124.3,172.4h24.5c34.9,0,42.9-26.5,42.9-39.7c0-21.5-13.7-39.7-43.7-39.7h-23.7V172.4z}
 svg{M88.7,56.8c0,5.5-4.5,10.1-10.1,10.1c-5.6,0-10.1-4.6-10.1-10.1c0-5.6,4.5-10.1,10.1-10.1C84.2,46.7,88.7,51.3,88.7,56.8z};
}}
\newcommand\orcidicon[1]{\href{https://orcid.org/#1}{\mbox{\scalerel*{
\begin{tikzpicture}[yscale=-1,transform shape]
\pic{orcidlogo};
\end{tikzpicture}
}{|}}}}

\newcommand{\aref}[1]{\hyperref[#1]{Appendix~\ref{#1}}}

\defcitealias{2018MNRAS.477.2716K}{KBFC18}

\usepackage[normalem]{ulem}

\definecolor{darkgreen}{rgb}{0.13, 0.55, 0.13}
\definecolor{brown}{rgb}{0.65, 0.16, 0.16}
\newcommand{\psnote}[1]{{\textcolor{brown}{[PS: #1]}}}

\title[RMHD PopIII simulations]{Population III star formation in the presence of turbulence, magnetic fields and ionizing radiation feedback}

\author[Sharda and Menon]{Piyush Sharda$^{\orcidicon{0000-0003-3347-7094}\,1}$\thanks{sharda@strw.leidenuniv.nl (PS)} and
Shyam H. Menon$^{\orcidicon{0000-0001-5944-291X}\,2,3}$\thanks{shyam.menon@rutgers.edu (SHM)}
\\
$^{1}$Leiden Observatory, Leiden University, P.O. Box 9513, 2300 RA Leiden, The Netherlands\\
$^{2}$Department of Physics and Astronomy, Rutgers University, 136 Frelinghuysen Road, Piscataway, NJ 08854, USA\\
$^{3}$Center for Computational Astrophysics, Flatiron Institute, 162 5th Avenue, New York, NY 10010, USA\\
}

\date{Accepted 2025 May 13. Received 2025 May 13; in original form 2024 May 23}

\pubyear{2024}

\begin{document}
\label{firstpage}
\pagerange{\pageref{firstpage}--\pageref{lastpage}}
\maketitle

\begin{abstract}
Turbulence, magnetic fields and radiation feedback are key components that shape the formation of stars, especially in the metal-free environments at high redshifts where Population III stars form. Yet no 3D numerical simulations exist that simultaneously take all of these into account. We present the first suite of radiation-magnetohydrodynamics (RMHD) simulations of Population III star formation using the adaptive mesh refinement (AMR) code FLASH as part of the POPSICLE project. We include both turbulent magnetic fields and ionizing radiation feedback coupled to primordial chemistry, and resolve the collapse of primordial clouds down to few au. We find that dynamically strong magnetic fields significantly slow down accretion onto protostars, while ionizing feedback, as expected, is largely unable to weaken gas accretion at early times. This is because the partially ionized \ion{H}{ii} region gets trapped near the star due to insufficient radiative outputs from the star. The maximum stellar mass in the HD and RHD simulations that only yield one star exceeds $100\,\rm{M_{\odot}}$ within the first $5000\,\rm{yr}$. However, in the corresponding MHD and RMHD runs, the maximum mass of Population III stars is only $60\,\rm{M_{\odot}}$. In other realizations where we observe widespread fragmentation leading to the formation of Population III star clusters, the maximum stellar mass is further reduced by a factor of few due to fragmentation-induced starvation. We thus show that magnetic fields are more important than ionizing feedback in regulating the mass of the star during the earliest stages of Population III star formation.
\end{abstract}




\section{Introduction}
\label{s:intro}
Population III stars formed out of metal-free gas in dark matter minihaloes in the early Universe. These stars were the building blocks of the first galaxies, initiated the enrichment of the Universe with dust and metals heavier than H, He and Li, and triggered the Epoch of Reionization. They are thought to have formed at very high redshifts ($ z> 15$), although recent works also point out the possibility of Population III star formation in metal-free pockets until the end of Epoch of Reionization at $ z \sim 6$ \citep{2018MNRAS.479.4544M,2020MNRAS.497.2839L,2023MNRAS.522.3809V}. Direct observations of metal-poor Population II stars in the Milky Way and nearby galaxies \citep[e.g.,][]{2011Natur.477...67C,2015MNRAS.453.2771J,2019MNRAS.488L.109N,2019ApJ...871..146F,2019ApJ...876...97E,2021ApJ...915L..30S,2022arXiv220803891M,2022A&A...668A..86A,2023A&A...669L...4A,2024arXiv240102484J}, supplemented by recent JWST discoveries of galaxies at $z > 10$ \citep[e.g.,][]{2023arXiv230600953M,2023MNRAS.525.4832Y,2023ApJS..265....5H,2023A&A...677A..88B,2023MNRAS.518.6011D,2023arXiv231104279F} highlight the importance of understanding how Population III stars formed and subsequently led to the formation of the first galaxies. However, no Population III stars have ever been directly observed, and current evidence or possible signatures remain weak, requiring theoretical developments to guide observations.

Numerous studies have been performed over the years to simulate the formation of the first stars \citep[e.g.,][]{2002ApJ...564...23B,2002Sci...295...93A,2012MNRAS.422..290S,2012MNRAS.424..399G,2012ApJ...745..154T,2014ApJ...781...60H,2016ApJ...824..119H,2020MNRAS.497..336S}. These studies start from cosmological initial conditions or isolated primordial clouds that eventually collapse due to gravity. During this collapse, the response of gas temperature to increasing density is fundamentally different in the absence of dust and metals, since the only available gas coolants (H$_2$ and HD) are not as efficient as dust and metals, rendering the gas hotter and prone to fragmentation over a larger range in density \citep{1998ApJ...508..141O,2005ApJ...626..627O}. This means that simulations need to resolve a larger range in density to capture fragmentation as compared to Population I star formation at Solar metallicity. Additionally, the rate of change of chemical processes is at par with the freefall time of gas under collapse, meaning that chemistry is largely out of equilibrium during Population III star formation \citep{1998A&A...335..403G,2013ARA&A..51..163G}. Thus, it has become common practice to solve for primordial chemistry \textit{on the fly} with hydrodynamics (HD) while simulating the formation of the first stars. Over the last two decades, numerical techniques have steadily improved, and latest works routinely include and study the effects of non-equilibrium chemistry, turbulence, magnetic fields, and radiation feedback in Population III star formation \citep[e.g.,][]{2020MNRAS.494.1871W,2020MNRAS.497..336S,2021MNRAS.505.4197S,2022MNRAS.516.3130S,2022MNRAS.512..116J,2022MNRAS.516.2223P,2023ApJ...959...17S}. 

Several papers have demonstrated the importance of interstellar turbulence, magnetic fields and radiation feedback in (massive) star formation in general \citep[e.g.,][]{2011ApJ...742L...9C,2012ApJ...761..156F,2012ApJ...754...71K,2016MNRAS.460.3272K,2016ApJ...832...40K,Guszejnov16a,2020AJ....160...78R,2021MNRAS.507.2448M,2020MNRAS.493.4643M,2021MNRAS.500.1721M,2022MNRAS.512..216G,2023MNRAS.522.5374H}, and for Population III star formation in particular \citep{2013MNRAS.432..668L,2016MNRAS.462.1307S,2022MNRAS.511.5042S,2016ApJ...824..119H,2020ApJ...892L..14S,2023ApJ...959...17S,2020MNRAS.497..336S,2021MNRAS.503.2014S,2021MNRAS.505.4197S,2022ApJ...935L..16H}. The latter set of papers conclude that both magnetic fields and radiation feedback are important during Population III star formation, and akin to Population I, play a crucial role in setting the Population III initial mass function (IMF, \citealt{2019FrASS...6....7K,2023ARA&A..61...65K,2024arXiv240407301H}).

However, none of the existing works on Population III star formation have simultaneously included turbulence, magnetic fields and radiation feedback. Therefore, the extent of their relative importance remains unclear. Understanding the physics regulating Population III stellar masses is crucial to address the uncertainty surrounding the potential upper mass limit for Population III stars, with hypotheses extending to stars as massive as $10^5\,\rm{M_{\odot}}$ \citep[e.g.,][]{2016ApJ...830L..34U,2016A&A...585A..65H,2018MNRAS.474.2757H}, which have been suggested to act as seeds for supermassive black holes \citep[see reviews by][]{2019PASA...36...27W,2020ARA&A..58...27I}. Similarly, a thorough comparison of Population III versus Population I star formation has not yet been possible, so it is also not clear how star formation proceeds and IMF varies as a function of interstellar medium (ISM) metallicity (see, however, \citealt{2021MNRAS.508.4175C,2023arXiv231213339C,2022MNRAS.509.1959S,2023MNRAS.518.3985S} for recent progress in this area). Part of this is due to computational feasibility of running these simulations long enough for stars to reach the zero age main sequence (ZAMS), but also due to the fact that radiation magneto-hydrodynamics (RMHD) needs to be coupled to chemistry (that becomes increasingly non-equilibrium in densest regions as the metallicity decreases -- \citealt{2005ApJ...626..627O,2009MNRAS.393..911G}) to self-consistently evolve radiation and chemical parameters in simulations \citep{2018MNRAS.479.3206N,2020A&A...636A..68D,2022MNRAS.512..348K,2023ApJS..264...10K,MS24}. 

Motivated by this critical gap in our understanding of Population III star formation, we perform the first set of 3D RMHD simulations of Population III star formation with non-equilibrium on the fly chemistry, turbulence, magnetic fields and ionizing radiation feedback, as part of the POPSICLE project (Population II/III Simulations Including Chemistry, Luminosity and Electromagnetism). We arrange the rest of the paper as follows: \autoref{s:setup} presents the setup we use to develop and run the simulations, \autoref{s:results} presents the results for simulations where an isolated star forms, and \autoref{s:results_cluster} expands it to star clusters. Finally, we present our limitations and conclusions in \autoref{s:caveats} and \autoref{s:conclusions}, respectively.

\begin{table}
\centering
\caption{Description of physical processes included in the simulations used in this work. We conduct three separate turbulent realizations for the RHD and RMHD runs.}
\begin{tabular}{|l|c|c|c|c|}
\hline
Simulation & Primordial & Turbulence & Magnetic & Ionizing \\
& Chemistry & & Fields & Feedback \\
\hline
HD (Control run)  & \checkmark & \checkmark & $\times$ & $\times$ \\
MHD & \checkmark & \checkmark & \checkmark & $\times$\\
RHD & \checkmark & \checkmark & $\times$ & \checkmark \\
RMHD & \checkmark & \checkmark & \checkmark & \checkmark \\
\hline
\label{tab:tab1}
\end{tabular} \\
\end{table}

\section{POPSICLE Simulations}
\label{s:setup}
The POPSICLE project will be introduced in a companion paper (Sharda et al. in prep.); here, we highlight salient features relevant to this work. To distinguish the effect of different physical mechanisms on Population III star formation, we conduct four sets of simulations: HD, MHD, RHD, and RMHD. As the naming suggests, HD refers to purely hydrodynamic simulations, without magnetic fields and radiation feedback. MHD includes magnetic fields but not ionizing feedback, whereas RHD includes ionizing feedback but not magnetic fields. Lastly, RMHD includes both magnetic fields and ionizing feedback. We summarize this information in \autoref{tab:tab1}. By ionizing feedback, we refer to ionization of H and H$_2$ due to extreme-UV (EUV) photons with energies upwards of $13.6\,\rm{eV}$. We exclude the dissociation of H$_2$ due to far-UV (FUV) photons in the Lyman-Werner band in this work.

\subsection{Code description}
We use the adaptive mesh refinement (AMR) code FLASH \citep{2000ApJS..131..273F,2008ASPC..385..145D} to run the suite of 3D HD, MHD, RHD, and RMHD simulations. The simulations use a novel, non-local Variable Eddington tensor (VET) based radiative hydrodynamics scheme coupled to a primordial chemistry network \citep{MS24}, which we further describe in \autoref{s:setup_vettam}. The primordial chemistry network includes $\rm{H,\,H^+,\,H^-,\,H_2,\,H^+_2,\,D,\,D^+,\,D^-,\,HD,\,HD^+,\,He,\,He^+,\,He^{++}}$, and electrons, and is derived from the astrochemistry package \texttt{KROME} \citep{2014MNRAS.439.2386G}. We use the Bouchut solver for compressible MHD \citep{Bouchut2007,Bouchut2010}, adapted for FLASH by \citet{2009JCoPh.228.8609W,2011JCoPh.230.3331W}, and the Barnes Hut tree solver to model gas self-gravity and gravitational forces between the gas and sink particles \citep{2018MNRAS.475.3393W}. The gas temperature is controlled by the coupled evolution of the non-equilibrium chemistry network along with various heating/cooling processes -- cooling due to H$_2$ and HD, compressional heating, cooling due to recombinations and collisions, heating/cooling due to chemical reactions, compton cooling, and heating due to ionizing feedback. We follow \cite{2019MNRAS.490..513S} to calculate the adiabatic index of gas -- including the full rovibrational formalism for the adiabatic index of H$_2$ that varies as a function of density and temperature. 

\subsection{Initial conditions and refinement}
Our initial conditions are identical to previous 3D MHD simulations conducted using FLASH by \cite{2020MNRAS.497..336S,2021MNRAS.503.2014S}. Briefly, we begin with a spherical primordial cloud core of radius $1\,\rm{pc}$ embedded in a computational domain of size $2.4\,\rm{pc}$. The cloud has a mass $M_{\rm{cl}} = 1000\,\rm{M_{\odot}}$, and uniform density ($n = 9050\,\rm{cm^{-3}}$). We derive the initial cloud temperature ($T = 265\,\rm{K}$), mass fractions $x_{\mathrm{H}} = 0.7502,\,x_{\mathrm{H_2}} = 0.0006,\,x_{\mathrm{He}} = 0.2492$, and ionization fraction $y_{\mathrm{e^-}} = 8.85\times10^{-7}$ from one zone primordial chemistry models \citep[e.g.,][]{2005ApJ...626..627O,2014MNRAS.439.2386G} appropriate for this density. We also impose solid body rotation on the cloud along the $\hat{z}$ plane, with the initial rotational energy around 3 per cent of the gravitational energy \citep[e.g.,][]{2019MNRAS.490..513S,2020MNRAS.497..336S}. We drive trans-sonic turbulence initially, which follows a velocity power spectrum $P_{\mathrm{v}} \propto k^{-1.8}$, where the wave number $k$ spans the range $2 \leq k \leq 20$, with a natural mixture of solenoidal and compressive modes \citep{2010A&A...512A..81F,2011PhRvL.107k4504F}, using the method implemented in \citet{2022ascl.soft04001F} for these purposes. These initial conditions appropriately represent the onset of Population III star formation in dark matter minihaloes at $z \approx 30$ \citep[e.g.,][]{2002Sci...295...93A,2002ApJ...564...23B,2011Sci...331.1040C,2014ApJ...781...60H,2014ApJ...792...32S,2014ApJ...785...73S,2021ApJ...917...40K}.

We allow 10 levels of grid refinement based on the Jeans length, such that at the highest resolution the cell size is $7.5\,\rm{au}$, equivalent to a maximum effective resolution of $65,536^3$ grid cells. We use 64 cells per Jeans length to refine the grid in order to not only hinder artificial fragmentation \citep{1997ApJ...489L.179T} and resolve dynamo amplification of magnetic fields \citep{2011ApJ...731...62F,2011PhRvL.107k4504F,2012MNRAS.423.3148S}, but also to better resolve shock fronts and capture shock chemistry that has important consequences for the thermodynamic evolution of the gas during collapse \citep{2021MNRAS.503.2014S}.

\begin{figure}
\includegraphics[width=\columnwidth]{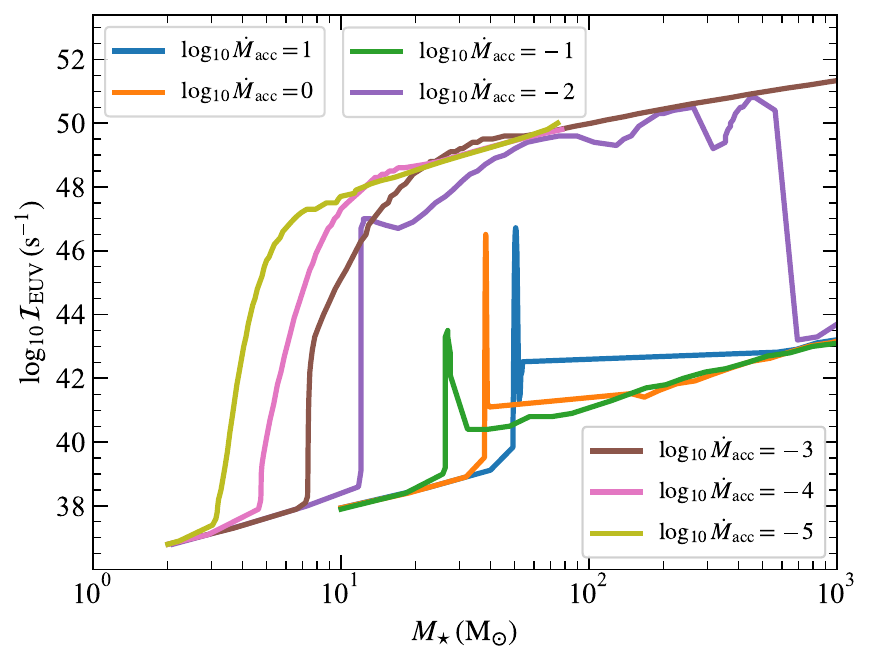}
\caption{Ionizing rate (in per s) of EUV photons in the energy range $13.6\,\rm{eV} - \infty$ as a function of the protostellar mass, for different mass accretion rates in the \texttt{GENEVA} Population III stellar evolution models used in this work \citep{2018MNRAS.474.2757H}. Lower accretion rates lead to higher effective temperatures, thus producing more ionizing photons.}
\label{fig:ieuv}
\end{figure}

We insert a sink particle at the maximum level of refinement once the density exceeds $10^{13}\,\rm{cm^{-3}}$; the sink particle formation criteria is described in detail in \cite{2010ApJ...713..269F,2011IAUS..270..425F}. The sink particle can move around the grid, and accrete mass from its surroundings, with the sink accretion radius set to $r_{\rm{sink}} = 2.5 \Delta x_{\rm{min}}$, where $\Delta x_{\mathrm{min}}$ is the minimum cell size at the highest level of refinement, to avoid artificial fragmentation very close to an existing sink.

\subsection{Magnetic fields}
It is now well known that magnetic fields can be exponentially amplified in turbulent environments due to the action of turbulent dynamo \citep[see the review by][]{2019Galax...7...47S}. However, reaching numerical convergence while resolving the action of turbulent dynamos is not currently possible given the high Reynolds numbers in the ISM. Nonetheless, previous works find that as long as the collapsing cloud is resolved with at least 32 cells per Jeans length, dynamo amplification of magnetic fields can be captured \citep{2011ApJ...731...62F,2012ApJ...745..154T,2012PhRvE..86f6412S}. 

Numerous works point out that it is plausible for a seed field to be present during the formation of the first stars \citep[][and references therein]{2020MNRAS.497..336S}. Following the results of \cite{2021MNRAS.503.2014S} who find that even an initially week seed field is quickly amplified to saturation due to the action of small-scale turbulent dynamo, we set our initial magnetic field strength to $30\,\mathrm{\mu G}$, which corresponds to a saturation state where the magnetic energy is 10 per cent of the turbulent kinetic energy \citep{2011PhRvL.107k4504F,2014ApJ...797L..19F,2016JPlPh..82f5301F,2015PhRvE..92b3010S}, appropriate for trans-sonic Mach numbers and near-unity Prandtl numbers as expected in the early Universe \citep{2004PhRvE..70a6308H,2005PhR...417....1B}.\footnote{Magnetic field strength of $30\,\mathrm{\mu G}$ corresponds to the B10 set of simulations run by \cite{2020MNRAS.497..336S,2021MNRAS.503.2014S}.} The power spectrum of magnetic fields goes as $P_{\mathrm{mag}} \propto k^{1.5}$ for $2 \leq k \leq 20$. The initial seed field is completely turbulent, since the field is not expected to be ordered prior to the formation of the first stars in the minihalo.

\subsection{Protostellar evolution}
To model the evolution of stellar radius, effective temperature, and luminosity in the protostellar phase, we use the stellar evolution models of \cite{2018MNRAS.474.2757H}. The models are calculated using a modified version of the 1D \texttt{GENEVA} stellar evolution code \citep{2008Ap&SS.316...43E} that solves for the structure of the stellar interior by taking the protostellar accretion rate into account \citep{2013A&A...557A.112H,2016A&A...585A..65H}. The \texttt{GENEVA} grid spans across a wide range of protostellar masses ($2 - 10^4\,\rm{M_{\odot}}$) and accretion rates ($10^{-5} - 10\,\rm{M_{\odot}\,yr^{-1}}$) that we interpolate across to obtain the stellar radius ($R_{\star}$) and effective temperature ($T_{\rm{eff}}$). If the accretion rate falls outside the range considered in the \texttt{GENEVA} model grids, we use the closest value of the stellar properties of interest that are available in the model grid \citep[e.g.,][]{2022MNRAS.512..116J}. We then use these quantities to estimate the radiative outputs from our stars. 

\subsection{Ionizing radiation feedback}
\label{s:setup_vettam}
To model the radiation transport of ionizing photons from our star particles we use the \texttt{VETTAM}\footnote{Variable Eddington Tensor-closed Transport on Adaptive Meshes} radiation (magneto-)hydrodynamics (RMHD) scheme \citep{2022MNRAS.512..401M}, and fully couple the photoionization to our primordial chemistry network; the details of this numerical approach (and tests) will be presented in a forthcoming paper \citep{MS24}; we only mention the salient features relevant to our study here. \texttt{VETTAM} uses the Variable Eddington Tensor closure by computing the Eddington tensor with a non-local ray-tracing approach, and uses this to close the system of RMHD equations; this non-local approach permits significantly higher accuracy in the radiation transport than methods that use local closure relations -- such as flux limited diffusion (FLD) or the Moment-1 (M1) closure -- especially in the presence of multiple radiation sources as in a star cluster \citep[see, for example, Fig. 13 in][]{2022MNRAS.512..401M}. To the best of our knowledge, this is the only suite of RMHD simulations that use VET based radiation hydrodynamics and is fully coupled to chemistry.

\texttt{VETTAM} uses an implicit global temporal update for the radiation moment equations for each band, including the radiative output from sink particles as a smoothed Gaussian source term in the moment equations that has a form 
\begin{equation}
    \label{eq:jstargauss}
    \mathcal{J}_{*}(r)=\frac{L_{*}}{\left(2 \pi \sigma_{*}^{2}\right)^{3 / 2}} \exp \left(-\frac{r^{2}}{2 \sigma_{*}^{2}}\right), 
\end{equation}
where  $L_{*}$ is the luminosity in the given energy band, $r$ is the radial distance of a grid cell from the sink particle, and $\sigma_* = 2 \Delta x_{\mathrm{min}}$.\footnote{Since $\sigma_{\star} < r_{\rm{sink}}$, this implies that we inject photons on the grid at scales smaller than our sink accretion radius (\textit{i.e.,} $\lesssim 20 \, \mathrm{au}$) and not at the edge of the sink accretion radius \citep[e.g.,][]{2014ApJ...792...32S,2016MNRAS.462.1307S,2016ApJ...824..119H}. \citet{2022MNRAS.512..116J} show that injecting the radiation in the vicinity of the star can lead to dynamically different outcomes of the launched ionization fronts; that being said, our resolution is slightly lower than that of \citeauthor{2022MNRAS.512..116J}, and likely larger than the disc scale height close to the star (the relevant scale to capture this effect) so it is unlikely our approach offers any advantage here. However, higher resolution runs in the future would certainly benefit from this approach.} We note that this numerical choice injects radiation onto the grid smoother than a point source; this would result in smoother radial profiles for the photoionization and heating rate than a point source, although the total energy and/or number of photons is conserved. This is a necessary compromise since we use an implicit moment-based formulation for the RMHD equations, and an excessively small value of $\sigma_*$ leads to convergence issues. 

To compute $L_{*}$, we obtain (interpolated) values for the stellar radius ($R_*$) and temperature ($T_*$) from our (subgrid) stellar evolution model using its instantaneous mass and accretion rates from the simulation, and assume the star radiates as a blackbody. This permits us to calculate $L_*$ as 
\begin{equation}
    L_{*} = 4\pi R_{*}^2 \int_{\nu_{\mathrm{min}}}^{\nu_{\mathrm{max}}}  B_\nu(T_{*}) d\nu,
\end{equation}
where $\nu_{\mathrm{min}}$ and $\nu_{\mathrm{max}}$ are the frequency limits of the two radiation bands that we evolve: one from $13.6\,\rm{eV} - 15.2\,\rm{eV}$ that can only ionize H, and another from $15.2\,\rm{eV} - \infty$ that can ionize both H and H$_2$. We can then estimate the rate of ionizing photons, $\mathcal{I}_{\nu}$, from the star as
\begin{equation}
    \mathcal{I}_{\nu} = 4\pi R_{*}^2 \int_{\nu_{\rm{min}}}^{\nu_{\rm{max}}}  \frac{B_\nu(T_{*})}{h \nu} d\nu\,.
\end{equation}
We will refer to $\mathcal{I}_{\nu}$ for the energy band ionizing H (energy range $13.6\,\rm{eV} - \infty$) as $\mathcal{I}_{\rm{EUV}}$. \autoref{fig:ieuv} plots $\mathcal{I}_{\rm{EUV}}$ as a function of the protostellar mass for different accretion rates, $\dot M_{\rm{acc}}$. Higher $\dot M_{\rm{acc}}$ makes the star bloat up, which reduces the effective temperature and brings down $\mathcal{I}_{\rm{EUV}}$. All the tracks plotted in \autoref{fig:ieuv} are almost identical to those from \citet{2009ApJ...691..823H} and \citet{2010ApJ...721..478H} using a different stellar evolution code, except for the one for $\dot M_{\rm{acc}} = 0.01\,\rm{M_{\odot}\,yr^{-1}}$, around which stellar evolution becomes sensitive to atmospheric parameters \citep{2023MNRAS.521..463H}. We refer the reader to \citet[section 4.3]{2018MNRAS.474.2757H} for a discussion on the possible sources of discrepancy between the two.

For each radiation band, we perform the radiation transport independently, and compute the photoionization rate using the absorbed radiation energy density in a cell, which we pass on to \texttt{KROME} for the photochemical reactions
\begin{equation}
\begin{aligned}
\mathrm{H}+\gamma & \rightarrow \mathrm{H}^{+}+\mathrm{e}^{-} \\
\mathrm{H}_2+\gamma & \rightarrow \mathrm{H}+\mathrm{H},
\end{aligned}
\end{equation}
Note that for $\mathrm{H}_2$, we make the assumption that the photoionized product, ${\mathrm{H}^{+}_2}$, rapidly undergoes dissociative recombination (due to its high reaction rate in dense ionized gas), resulting in the production of two hydrogen atoms. This quasi-dissociation approach for the outcome of $\mathrm{H}_2$ ionization has also been invoked in other works \citep[e.g.,][]{2022MNRAS.512..116J}. We also incorporate heating due to photoionization of these species in the thermal balance computed by \texttt{KROME}. The thermal energy deposited in the gas per photoionization is computed from the excess energy available per photoionization event, which depends on the stellar spectrum as follows
\begin{equation}
    \mathcal{E}_{\nu} = \frac{\int_{\nu_{\mathrm{min}}}^{\nu_{\mathrm{max}}} \left(B_\nu(T_*) \sigma_\nu h(\nu-\nu_0)/h\nu \right) \,  d\nu}{\int_{\nu_{\rm{min}}}^{\nu_{\rm{max}}} \left(B_\nu (T_*)  \sigma_\nu/h\nu \right) \, d\nu}.
\end{equation}
We will refer to $\mathcal{E}_{\nu}$ for the energy band ionizing H as $\mathcal{E}_{\rm{EUV,H}}$ and that ionizing H$_2$ as $\mathcal{E}_{\rm{EUV,H_2}}$. In the presence of multiple stars, we use the luminosity-weighted average of $\mathcal{E}_{\nu}$ over all the stars.  Following \citet{1989agna.book.....O}, we specify the frequency-dependent cross section of H as
\begin{equation}
\sigma_{\rm{H}} = 6.3 \times 10^{-18} \left(\frac{\nu}{13.6 \rm{eV}/h} \right)^{-3}\,.
\label{eq:sigmaH}
\end{equation}
We follow \citet{baczynski2015} to express the frequency-dependent cross section of H$_2$ ($\sigma_{\rm{H_2}}$) as a step function between $15.2 - 18.1\,\rm{eV}$, and assume it falls off as $\nu^{-3}$ for higher energies. This formulation is based on analytical work by \citet{2012JPhB...45i5203L}, and is consistent with experimental results to within 20 per cent \citep{1993JChPh..99..885C}.

We also include the effect of the radiation pressure on gas by depositing the momentum carried by ionizing photons that are absorbed. We invoke on-the-spot approximation for photons emitted via recombination of H$^+$ to the ground state, assuming that they are reabsorbed before they can travel far from their production sites \citep[e.g.,][]{1938ApJ....88...52B,1989agna.book.....O}.

We do not track the ionization of other species (like D, HD and He) due to stellar photons since He remains relatively inert and the mass fractions of D and HD are orders of magnitude less than H and H$_2$. We also do not include the effects of far-ultraviolet Lyman-Werner (LW) radiation emitted by stars which can photodissociate H$_2$ \citep[e.g.,][]{2011Sci...334.1250H,baczynski2015,2022MNRAS.512..116J,2023ApJ...959...17S}; earlier studies have shown that the effects of this is rather minor \citep{2016ApJ...824..119H,2022MNRAS.512..116J}. That being said, some studies suggest that the heating from photodissociation could affect the accretion rates onto stars \citep{2014ApJ...792...32S,2016MNRAS.462.1307S}. We will present calculations with the LW band included in an upcoming paper.

\subsection{Multiple turbulent realizations}
The lack of efficient cooling agents (\textit{i.e., } metals and dust) render Population III star formation more prone to fragmentation over a larger range in density as compared to Population I star formation \citep[e.g.,][]{2023arXiv231010730P}. This is further complicated by the fact that the outcomes of star formation in cloud cores are stochastic in nature due to the presence of turbulent fluctuations \citep[][figure 7]{2020MNRAS.497..336S}. Therefore, it is insufficient to run a single suite of simulations with a particular random seed and conclude a statistically significant result \citep{2004A&A...414..633G,2004A&A...423..169G,2020MNRAS.494.1871W}. Further, \cite{2020MNRAS.497..336S} highlight how the impact of magnetic fields on the Population III IMF cannot be ascertained by running only a few simulations, since the initial random fluctuations caused by turbulence can significantly alter the resulting stellar masses from one simulation to the next. This complication is not limited to star formation -- recent works on cosmological simulations also find that stochasticity leads to significant scatter in the outcome for identical initial conditions \citep{2022MNRAS.515.1430D,2024MNRAS.527.4705D,2023MNRAS.526.2441B}.

To balance the stochasticity of outcomes and computational feasibility, we conduct 3 turbulent realizations for each subcategory of simulations (HD, MHD, RHD and RMHD), where each realization uses the same set of (pseudo-)random seeds to generate the initial turbulent velocity and magnetic fields. These realizations can be considered as representing the innermost regions of different dark matter minihaloes with varying potentials and density fields. Three realizations are sufficient for our purposes since our aim here is not to derive an IMF (that would otherwise require $\gtrsim 12$ realizations). In \autoref{s:results}, we first present and discuss the results for a set of simulations that \textit{mostly} formed only one star by the time we stop the simulations. This particular set of simulations is easier to analyze since gas dynamics around the star are not impacted by fragmentation. In \autoref{s:results_cluster}, we then describe the evolution of the other two realizations of the RHD and RMHD simulations where we find widespread fragmentation and formation of Population III star clusters.

\begin{figure}
\includegraphics[width=\columnwidth]{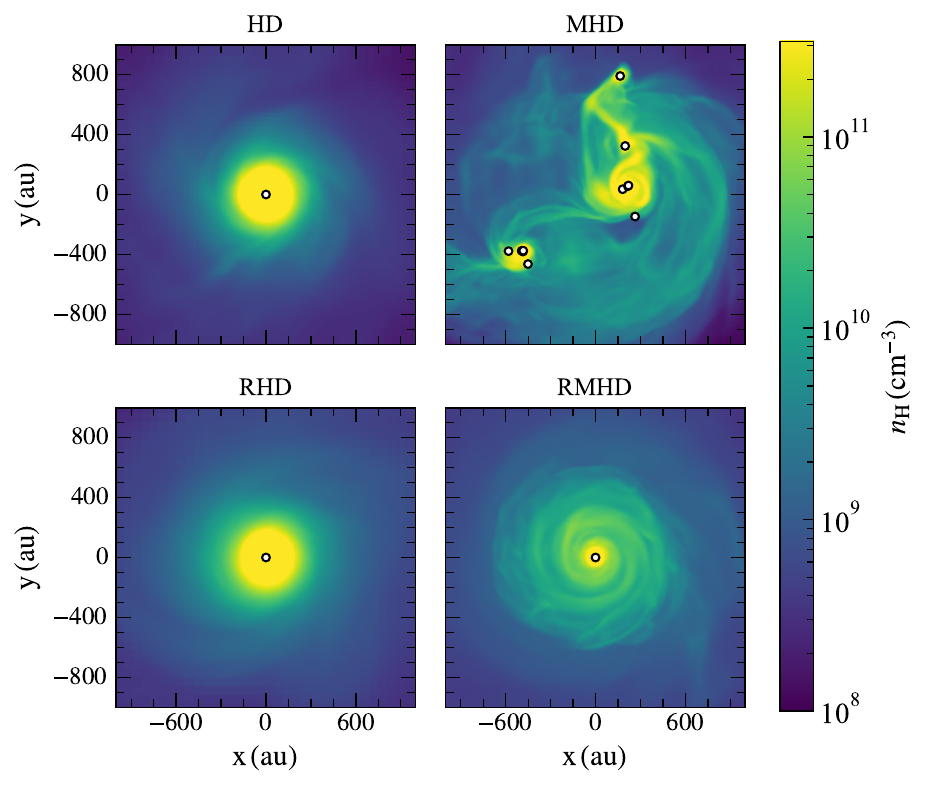}
\caption{Face-on density-weighted projections of the gas number density through the $\hat{z}$ axis of the simulation box at the end of the simulations. The four panels correspond to the runs with hydrodynamics (HD), magneto-hydrodynamics (MHD), radiation-hydrodynamics that includes ionizing radiation feedback (RHD), and radiation-magnetohydrodynamics (RMHD). White dots represent the position(s) of sink particle(s), used as a proxy for Population III stars. All except the MHD run only produces a single star. The accretion disc around the star in the RMHD simulation is less dense as compared to the other simulations. The mass of the central star is $120\,\rm{M_{\odot}}$ (HD), $60\,\rm{M_{\odot}}$ (MHD), $110\,\rm{M_{\odot}}$ (RHD), and $57\,\rm{M_{\odot}}$ (RMHD).}
\label{fig:proj_numdens}
\end{figure}

\begin{figure}
\includegraphics[width=\columnwidth]{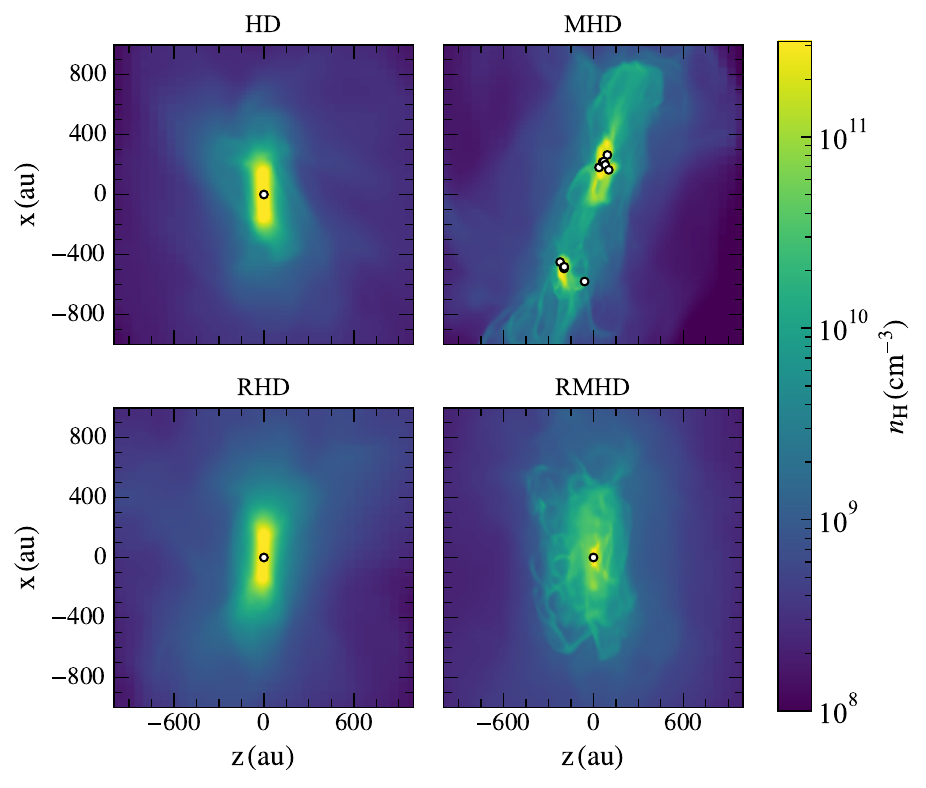}
\caption{Same as \autoref{fig:proj_numdens} but viewed edge-on along the $\hat{y}$ axis.}
\label{fig:proj_numdens_edgeon}
\end{figure}

\section{Results and discussion}
\label{s:results}
We evolve the simulations until 5000 yr post the formation of the first star. We first present and discuss results for the simplest turbulent realization, which only produces a single Population III star in the HD, RHD, and RMHD runs but shows fragmentation in the corresponding MHD run. This set of simulations is easy to follow because (except for the MHD case) it produces a single star with a well-defined accretion disc around it and makes the comparison between them straightforward. We first show and contrast the evolution of thermodynamic quantities (density, temperature, and mass fraction of H$_2$) in the four subsets of simulations, and then move on to the properties of magnetic fields and ionizing feedback and how their interplay governs gas fragmentation, star formation and accretion. Movies associated with the evolution of various parameters of interest are available as supplementary (online only) material.

\subsection{Physical and chemical properties}
\label{s:results_physical}
\autoref{fig:proj_numdens} plots the density-weighted projections of the number density of the gas within a $0.01\,\rm{pc}$ region centered on the star (center of all stellar masses for the MHD run), at the end of the simulations. We simply use the $\hat{z}$ axis of the simulation box as the axis through which we apply the projection. We see that the HD, RHD, and RMHD runs all yield a single massive star with a well-defined accretion disc around it. The disc is geometrically thin when viewed along the edge-on axes (see \autoref{fig:proj_numdens_edgeon}). It is also clear that the accretion disc is denser in the HD and RHD runs, as compared to the RMHD run. The MHD run fragments to produce a cluster of 12 stars.

\begin{figure}
\includegraphics[width=\columnwidth]{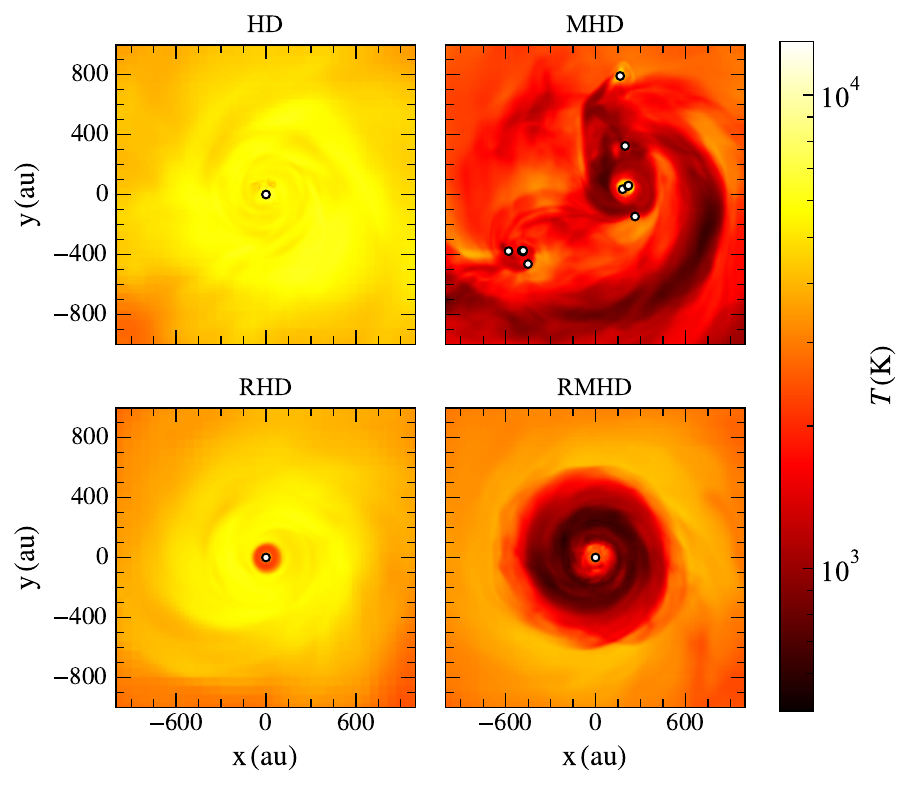}
\caption{Same as \autoref{fig:proj_numdens} but for the density-weighted gas temperature. The accretion disc is much colder in runs including magnetic fields (MHD and RMHD).}
\label{fig:proj_temp}
\end{figure}

The differences between the HD and the RHD runs are more prominent in the temperature structure around the star, as we read off from \autoref{fig:proj_temp}. In the HD run, gas rapidly falls onto the sink, creating shocks that cause heating and dissociation of H$_2$; we describe it further in the paragraph below. The RHD run contains an inner region that is colder than the HD run, which at first glance might seem counter-intuitive. Closer inspection reveals that this outcome is a result of radiation pressure acting on the gas in the immediate vicinity of the star that slows down the rate of (radial) gas infall, which weakens the shock produced by the inflowing gas (see \aref{s:app_radpres} for details). This is also why we observe enhanced levels of molecular hydrogen in the inner disc in the RHD run (see \autoref{fig:proj_h2}). 

In both the HD and RHD runs, the hot gas ($T > 5000\,\rm{K}$) all throughout the projection window is a consequence of the dependence of gas cooling on the Jeans resolution in the presence of accretion shocks created by matter falling onto the disc. As \citet[appendix A]{2021MNRAS.503.2014S} explain, if the Jeans length is well resolved, the dissociation timescale for molecular hydrogen in the post-shock region is smaller than the (resolved) gas cooling timescale, which ensures that H$_2$ dissociates faster than the gas can cool, thus giving rise to high gas temperatures. As we will explore below, this effect of the Jeans resolution on the chemistry is more crucial than protostellar radiation effects throughout the duration of the simulations. The high temperatures caused by shocks also enhance the mass fraction of H$^+$ in the HD and RHD runs, as we see from the projections of the mass fraction of H$^+$ we plot in \autoref{fig:proj_hp}. We observe a partially ionized \ion{H}{ii} region near the star, absent in runs without ionizing feedback, clearly indicating its origin in stellar feedback. The average mass fraction of H$^+$ within the region is $\sim 10^{-3}$, which implies that they are only marginally ionized, unlike fully ionized \ion{H}{ii} regions \citep[e.g.,][]{1998PASP..110..761F}, and steeply declines outside this region\footnote{We reiterate that our numerical choice of a Gaussian input for the stellar injection term (see Section~\ref{s:setup_vettam}) smoothens the radial variation of the ionization fraction than one would obtain from a point source, although the total number of ionizations are conserved.}. The \ion{H}{ii} region is unable to expand for the duration of these simulations. We will explore this further in \autoref{s:results_hii} and \autoref{s:results_cluster}.

Perhaps the most striking feature in the temperature maps is the presence of large amounts of cold gas around the star(s) in runs including magnetic fields (MHD and RMHD). The RMHD run is slightly different from the MHD run in that the former contains hot gas in the immediate vicinity of the star because of ionizing feedback. This feature is also noticeable in earlier MHD simulations of Population III star formation \citep{2020MNRAS.497..336S,2021MNRAS.503.2014S,2022MNRAS.516.3130S,2022MNRAS.511.5042S}, and cosmological simulations that use high Jeans resolution \citep{2012ApJ...745..154T,2024A&A...684A.195D}. To distinguish between the role of magnetic fields and gas temperature in the MHD and RMHD runs, we plot projections of plasma $\beta$ (defined as the ratio of the thermal to magnetic pressure in the gas) in \autoref{fig:proj_plasmabeta}. We find that plasma $\beta < 2$ close to the stars where the gas pressure is low due to lower temperatures. The gas in these regions is therefore sub-Alfvénic, and pressure provided by magnetic fields acts against gravity, which in turn brings down the rate of compressional heating \citep[e.g.,][]{2022MNRAS.516.3130S}. Lower temperatures lead to a larger abundance of molecular H$_2$ in the MHD and RMHD runs, as shown in \autoref{fig:proj_h2}. The initial magnetic field is already saturated so we do not expect the small-scale turbulent dynamo to be operational and responsible for these effects. We verify that this is indeed the case in \aref{s:app_dynamo}.  The gas farther out from the star in the RMHD is cold -- as it is shielded from the stellar radiation -- but also less dense, so it cannot fragment. This partially explains why the MHD run fragments to form more than one star but the RMHD run does not (see also, \autoref{s:results_accrates}). However, we cannot comment on whether this effect is always present based on a single turbulent realization; in fact, as we will see in \autoref{s:results_cluster}, other RMHD realizations show widespread fragmentation.

\begin{figure}
\includegraphics[width=\columnwidth]{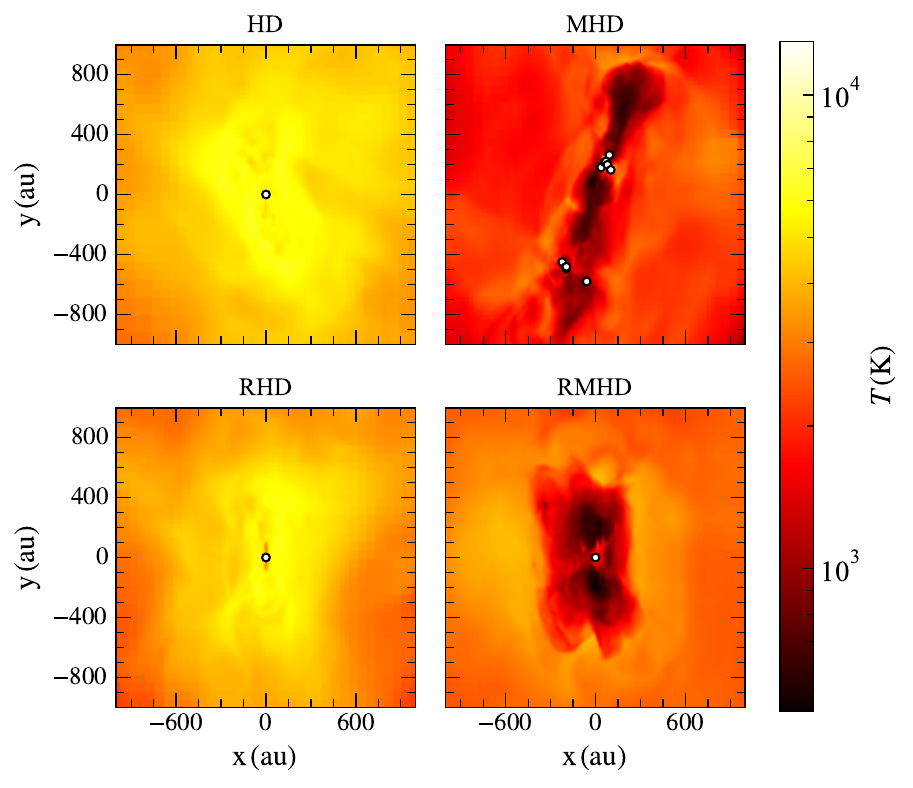}
\caption{Same as \autoref{fig:proj_temp} but viewed edge-on, along the $\hat{y}$ axis.}
\label{fig:proj_temp_edgeon}
\end{figure}

\begin{figure}
\includegraphics[width=\columnwidth]{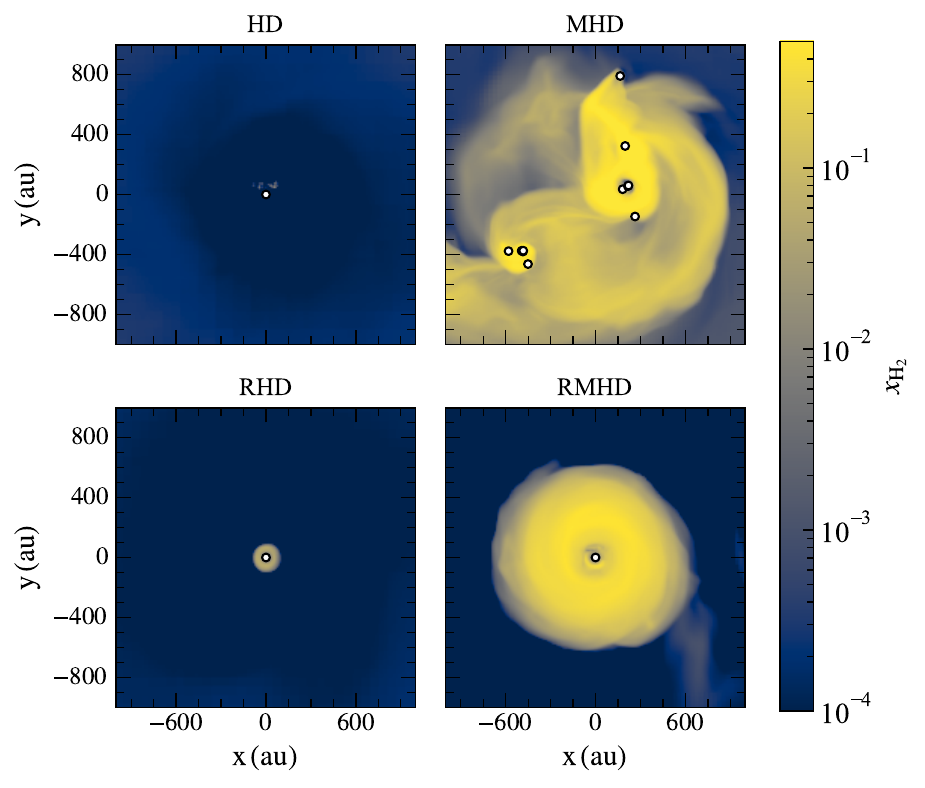}
\caption{Same as \autoref{fig:proj_numdens} but for the density-weighted mass fraction of molecular hydrogen. The accretion disc is highly molecular in the runs including magnetic fields (MHD and RMHD).}
\label{fig:proj_h2}
\end{figure}

\begin{figure}
\includegraphics[width=\columnwidth]{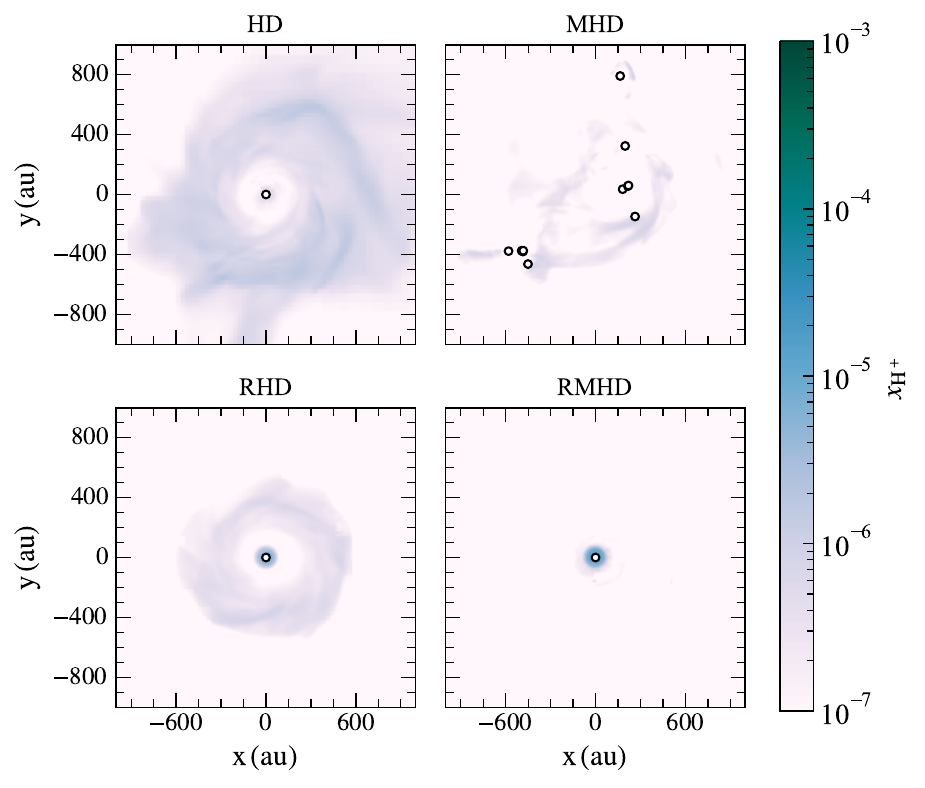}
\caption{Same as \autoref{fig:proj_numdens} but for the density-weighted mass fraction of H$^+$. The impact of ionizing feedback is noticeable close to the star in the RHD and RMHD runs, where $x_{\rm{H^+}}$ is higher. The large scale diffuse $x_{\rm{H^+}}$ seen in the HD and RHD runs is a result of hydrodynamic shocks.}
\label{fig:proj_hp}
\end{figure}

\begin{figure}
\includegraphics[width=\columnwidth]{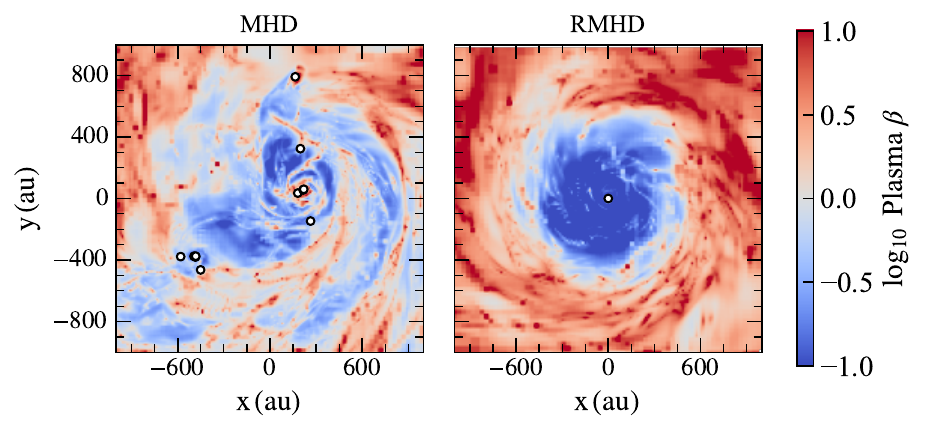}
\caption{Density-weighted projection of plasma $\beta$ for the same projection window as in \autoref{fig:proj_numdens}. Plasma $\beta$ denotes the ratio of the gas thermal to the magnetic pressure. Regions where the average plasma $\beta \gtrsim 2$ are super-Alfvénic, and those with $\beta \lesssim 2$ are sub-Alfvénic.}
\label{fig:proj_plasmabeta}
\end{figure}

\subsection{Radial and polar profiles}
\label{s:results_profiles}
\begin{figure*}
\includegraphics[width=\textwidth]{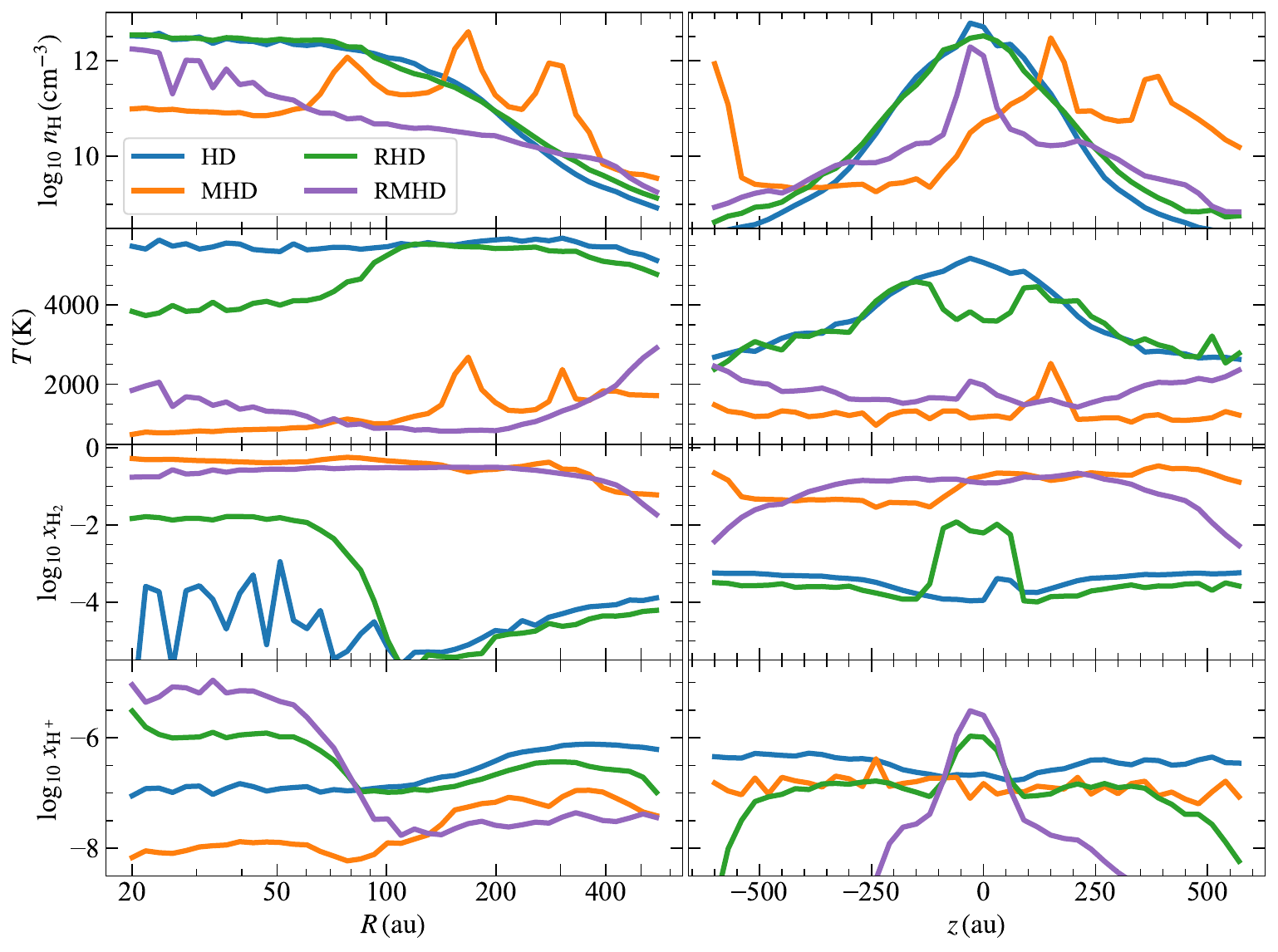}
\caption{\textit{Left panel:} mass-weighted radial profiles in the accretion disc of the gas number density, temperature, and mass fractions of H$_2$ and H$^+$ at the end of the simulations. The RMHD run exhibits lower temperatures, higher mass fractions of H$_2$ and lower densities as compared to the RHD run. Ionization of H is most noticeable close to the star where there is an overabundance of H$^+$ as compared to the control (HD) run. \textit{Right panel:} Same as the left panel but for profiles in the polar direction.}
\label{fig:radialprofiles}
\end{figure*}

We now look at the (cylindrical) radial and polar profiles of key physical parameters in the four simulations. \autoref{fig:radialprofiles} plots the mass-weighted radial and polar profiles of gas density, temperature, and mass fractions of H$_2$ and H$^+$ at the end of the simulations. For the purposes of this discussion, we classify the inner accretion disc as the region within $100\,\rm{au}$ of the star, and outer accretion disc as the region between $100-600\,\rm{au}$ from the star. \autoref{fig:radialprofiles} shows that the radial density profiles in the accretion disc are very similar between the HD and RHD runs -- a mild decline in the inner disc that then changes to a steeper decline in the outer disc. The disc in the RMHD run has more diffuse gas as compared to the HD and RHD runs, as also expected from \autoref{fig:proj_numdens}. On the other hand, the MHD run shows a complex radial density profile, with peaks that correspond to companion stars present within the disc of the most massive star. In the polar direction, the HD, RHD, and RMHD runs show a density profile that peaks near $z=0$ and smoothly declines in either direction, although the decline is steeper by a factor of a few in the polar direction.

The temperature profile in the inner disc is quite distinct in all the four runs, and are set by distinct physics. The high temperature near the star in the HD run is a consequence of coupled chemistry and cooling, which can occur due to the high Jeans resolution we achieve in our simulations, as we explain in \autoref{s:results_physical}. The temperature is lowest in the MHD run due to magnetic fields reducing compressional heating, low enough to trigger fragmentation and formation of multiple stars. The somewhat lower temperature in the RHD run compared to the HD run is due to the effects of radiation pressure, as we explain in \aref{s:app_radpres}. Finally, the temperature in the RMHD run in the inner disc is set by a mixture of all the three above: Jeans resolution, ionization -- recombination balance, and magnetic fields. Overall, the effect of magnetic fields seem to be the strongest, since the temperature in the RMHD run is closest to the MHD run.

As expected based on the temperature profiles and results in \autoref{s:results_physical}, the MHD and RMHD runs contain a significant amount of H$_2$ in the inner disc. The temperature closest to the star in the RMHD run is high enough to cause some H$_2$ dissociation, which slightly lowers the mass fraction of H$_2$ as compared to the MHD run. The RHD run also contains a non-negligible amount of H$_2$ in the inner disc, whereas the HD run has no discernible H$_2$. The findings are qualitatively similar in the polar direction as well: the MHD and RMHD runs have a high fraction of H$_2$, and the RHD run has non-negligible H$_2$ close to the star. The last panels of \autoref{fig:radialprofiles} clearly show an enhancement in the fraction of H$^+$ close to the star in the runs including ionizing feedback (RHD and RMHD). Nevertheless, the mass fraction of H$^+$ is negligible, as we explore further in \autoref{s:results_hii} below.

\autoref{fig:velocityprofiles} shows the toroidal and turbulent velocity profiles for the HD, RHD, and RMHD runs.\footnote{The turbulent component of the velocity is calculated by subtracting mean velocity in each radial bin, explained in detail in \citet[section 3.2]{2021MNRAS.503.2014S}.} The profiles are non-linear and less intuitive in the MHD simulation due to fragmentation, so we do not show it here. The accretion disc in the HD, RHD, and RMHD runs show sub-Keplerian profiles. The ordered and turbulent velocity components in the three runs are similar, but lower temperatures in the RMHD run lead to higher turbulent Mach numbers, possibly explaining the presence of density structures we see in \autoref{fig:proj_numdens}.

\begin{figure}
\includegraphics[width=\columnwidth]{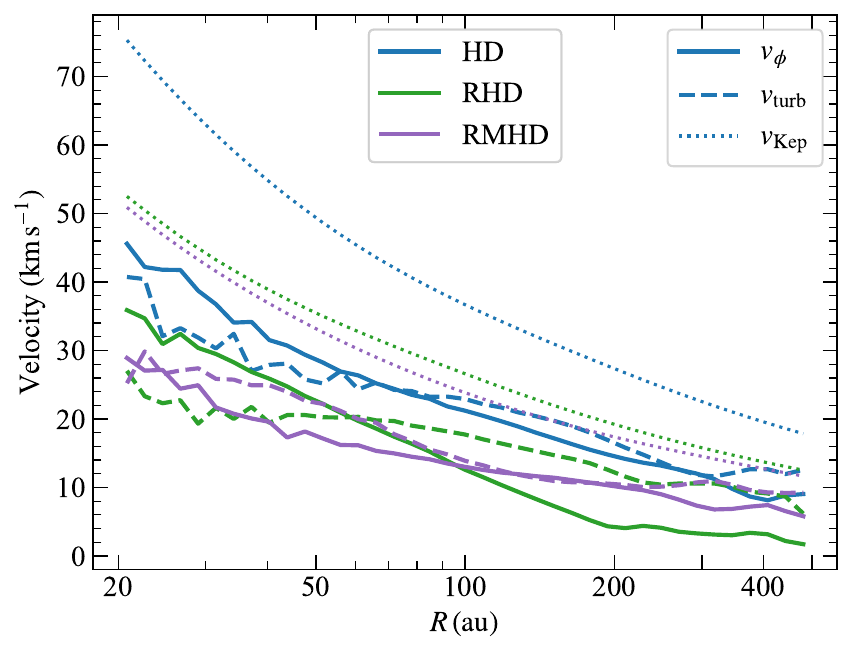}
\caption{Mass-weighted radial profiles of the toroidal ($v_{\phi}$, solid), and the turbulent ($v_{\rm{turb}}$, dashed) components of velocity in the HD, RHD and RMHD simulations. Overplotted in dotted curves is the Keplerian velocity ($v_{\rm{Kep}}$) in each case.}
\label{fig:velocityprofiles}
\end{figure}

\subsection[Evolution of Hii regions]{Evolution of \ion{H}{II} regions}
\label{s:results_hii}
We have seen in \autoref{s:results_physical} and \autoref{s:results_profiles} that the impact of radiation on the evolution of the clouds is quite modest. This is primarily because the stars are unable to ionize the dense gas in vicinity of stars, and only forms a partially ionized \ion{H}{ii} region. This can also be seen in the radial profiles of the mass fraction of H$^+$ (cf. \autoref{fig:radialprofiles}), which, while higher than the control run without radiation, are still quite low: $x_{\rm{H^+}} \lesssim 10^{-6}$. The weak effect of radiation can be understood by comparing the ionizing photon rates to the gas densities by computing the size of the Stromgren sphere that could be driven \citep{1939ApJ....89..526S}. If the Stromgren radius is much smaller than the cell size, it implies that the photon rate is insufficient to ionize the cells in the vicinity of the source, and can only partially do so. The Stromgren radius for an ionized hydrogen bubble under the on-the-spot approximation is given by 
\begin{equation}
    R_{\mathrm{St}} = \left( \frac{3\mathcal{I}_{\rm{EUV}}}{4 \pi \alpha_{\mathrm{B}}n_{\mathrm{H}}^2} \right)^{1/3} \sim 0.14 \, \mathrm{au} \, \left(\frac{\mathcal{I}_{\rm{EUV}}} {10^{49} \, \mathrm{s}^{-1}} \right)^{1/3} \left(\frac{n_{\mathrm{H}}}{10^{12} \, \mathrm{cm}^{-3}} \right)^{-2/3}
\end{equation}
where $n_{\mathrm{H}}$ the hydrogen number density and $\alpha_{\mathrm{B}}$ the case-B recombination coefficient; in the dimensional scaling above, we have used the values for $\mathcal{I}_{\rm{EUV}}$ and $n_{\mathrm{H}}$ that we find for our Population III star and its vicinity respectively, and the value of $\alpha_{\mathrm{B}} = 2.6\times10^{-13}\,\rm{cm^{3}\,s^{-1}}$ from \citet{1997MNRAS.292...27H} appropriate for a temperature of $10^4$K.\footnote{In reality, the gas temperature in the partially ionized regions in our simulations is less than $10^4\,\rm{K}$ by a factor $1.5-3$. The case B recombination coefficient is larger at lower temperatures \citep[][Appendix A]{1997MNRAS.292...27H}, so the expected $R_{\rm{St}}$ is slightly lower than that quoted in the main text.} The value reported above is smaller than the cell size of our simulations, explaining why we do not fully ionize cells close to the star. Note that this does not mean that if we had higher resolution then the star would have driven an ionized bubble -- the densities one would resolve would also be higher in this case, and the same problem would likely occur. This is rather a statement on the weak ionizing luminosities of the star (due to high accretion rates) for the densities with which it is surrounded. We also do not see any expansion of the \ion{H}{ii} region in the polar directions where the densities within $\pm 100\,\rm{au}$ are lower but the Stromgren radius remains unresolved. 

The calculation above implies that $\mathcal{I}_{\rm{EUV}}$ needs to increase and $n_{\mathrm{H}}$ needs to decrease for effectively driving a \ion{H}{ii} region. This is highly likely to occur at later times as the star becomes more massive and emits more ionizing photons, while $n_{\mathrm{H}}$ decreases as mass is increasingly depleted in the vicinity of the star -- especially in the polar directions. Another, more subtle effect could arise from the evolution of the mass accretion rate onto the star. In the absence of any gas infall onto the minihalo, the accretion rate onto the star would naturally lower with time as the disc/cloud gets increasingly depleted, and any fragmentation would decrease the accretion rate onto the primary star further due to competitive accretion. In fact, in agreement with our expectations, other RHD simulations of Population III star formation in the literature find that \ion{H}{ii} regions are driven at times $\gtrsim 15000 \, \mathrm{yr}$ \citep{2016ApJ...824..119H,2020ApJ...892L..14S}, while the effects at earlier times are modest. We will be well placed to comment on these effects in a forthcoming paper where we follow the evolution of the simulation for significantly longer times. 

We note that \citet{2022MNRAS.512..116J} report that \ion{H}{ii} regions in their simulations are trapped in the disc, and therefore do not have a significant effect on the system. We pause to clarify that the physical reason there is distinct from what is described above. In their work, they do ionize the cells in the close vicinity of the star since they 1.) evolve for much longer times, and 2.) have an effective grid resolution $\lesssim 1 \, \mathrm{au}$. However, this ionized bubble is unable to undergo a (D-type) pressure-driven expansion, even in the polar directions where densities are lower, and remains trapped within the disc -- unlike several other works in the past \citep[see Table 4 in][]{2022MNRAS.512..116J}. This occurs because the gravitational influence of the star is so strong such that the ionized gas is unable to accelerate and expand (see their figure 18); a scenario that is not only limited to Population III star formation \citep[][]{2002ApJ...580..980K,2018MNRAS.480.3468K,2019MNRAS.486.5171S}. We cannot comment on this possibility in this work since our ionizing luminosities at the times we evolve are too weak to even produce an ionized bubble (\textit{i.e.,} the Stromgren sphere) that we resolve.

\begin{figure}
\includegraphics[width=\columnwidth]{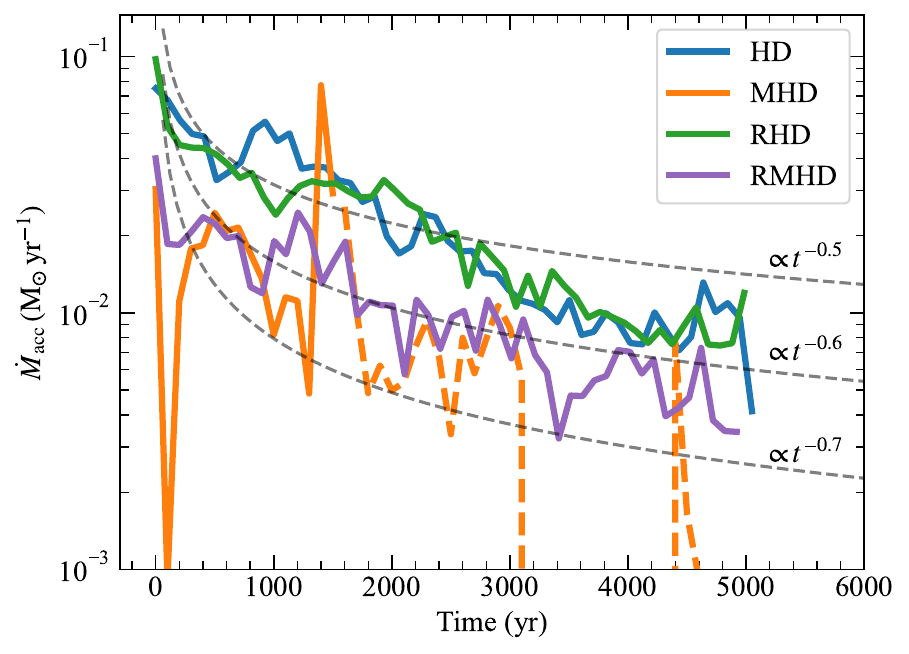}
\caption{Average mass accretion rates onto the protostar (in case of MHD, the oldest and most massive protostar) in the four simulations. Fragmentation in the MHD run occurs at $1500\,\rm{yr}$ after the formation of the first star, which is indicated by the onset of the dashed orange curve, to emphasize that the star does not evolve in isolation thereafter. Simulations including magnetic fields (MHD and RMHD) exhibit lower protostellar accretion rates. Dashed grey curves show example trends for $\dot M_{\rm{acc}} \propto t^{-\beta}$, with $\beta = 0.5, 0.6, 0.7$, as shown on the plot.}
\label{fig:accr}
\end{figure}

\begin{figure*}
\includegraphics[width=\textwidth]{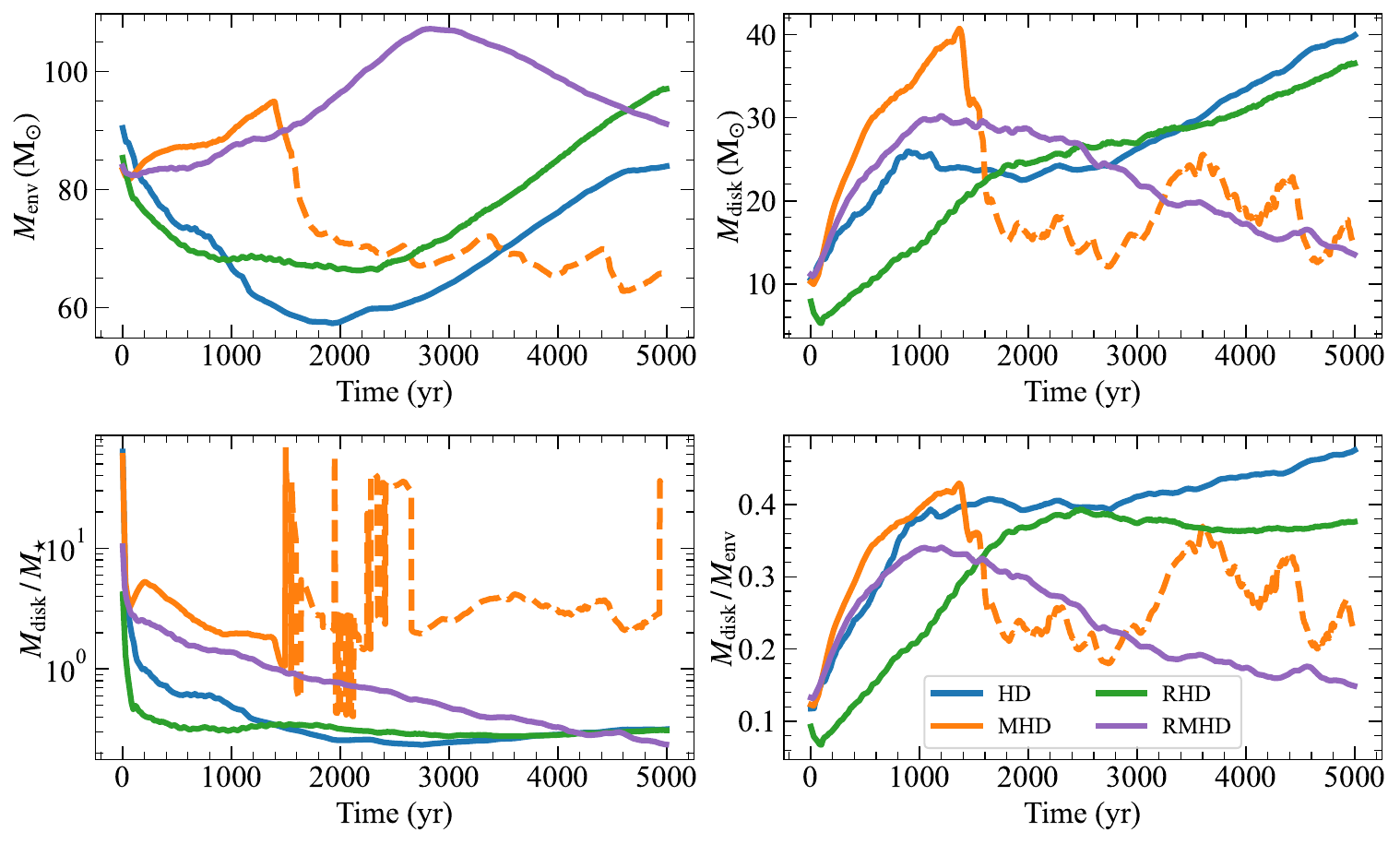}
\caption{Sequence of mass transport from the largest to smallest scales. \textit{Top left panel:} Evolution of mass enclosed within an envelope of radius $0.01\,\rm{pc}$ centered at the location of the star (in the HD, RHD, and RMHD simulations) and at the location of the most massive star in the MHD simulation. The MHD run fragments $1500\,\rm{yr}$ after the formation of the first star, shown by the dashed orange curve. \textit{Top right panel:} Same as the top left panel but for the accretion disc within $500\,\rm{au}$ of the star. \textit{Bottom right panel:} ratio of mass in the disc to that in the envelope. \textit{Bottom left panel:} ratio of mass in the accretion disc to the protostellar mass. At each successive stage (from larger scales to the protostar), magnetic fields slow down mass transport, ultimately leading to lower stellar masses in the MHD and RMHD simulations (cf. \autoref{fig:mass}).}
\label{fig:disks}
\end{figure*}

\begin{figure}
\includegraphics[width=\columnwidth]{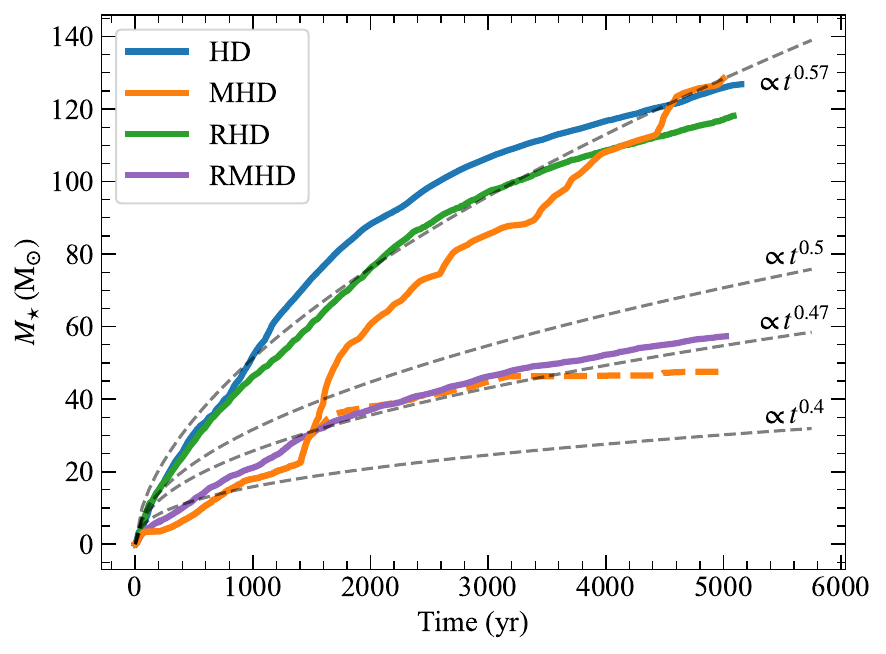}
\caption{Mass growth of the single protostar in the HD, RHD, and RMHD simulations. Fragmentation in the MHD run occurs at $1500\,\rm{yr}$ after the formation of the first star; this is indicated by the onset of the dashed orange curve to emphasize that the star does not evolve in isolation thereafter. The solid orange curve denotes the total mass accumulated in all the stars in the MHD run. Dashed grey curves demarcate example trends to guide the eye for the relation $M_{\star} \propto t^{\gamma}$ with $\gamma = 0.57, 0.5 0.47$ and $0.4$. Lower accretion rates (onto the most massive protostar) in the MHD and RMHD runs lead to lower stellar masses (by a factor $\gtrsim 2$) as compared to the HD and RHD runs.}
\label{fig:mass}
\end{figure}

\begin{figure*}
\includegraphics[width=\textwidth]{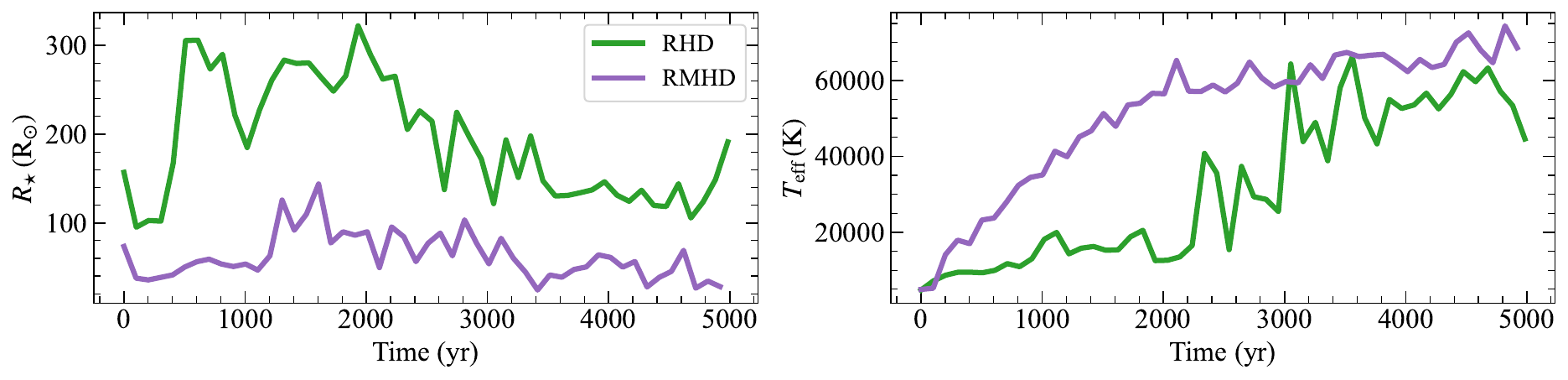}
\caption{Temporal evolution of stellar properties based on the \texttt{GENEVA} stellar structure models \citep{2018MNRAS.474.2757H}, using the instantaneous mass and accretion rates of the star, compared between the RHD and RMHD runs. \textit{Left panel:} stellar radius. \textit{Right panel:} effective temperature. The star in the RMHD run is more compact and bluer than that in the RHD run even though its mass is less by a factor of two.}
\label{fig:radius_teff}
\end{figure*}

\begin{figure*}
\includegraphics[width=\textwidth]{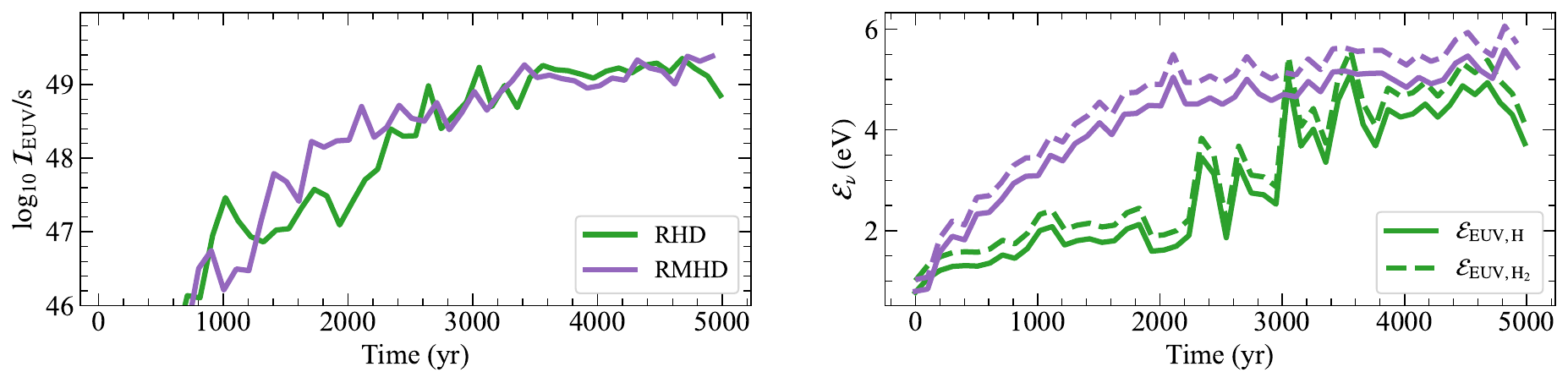}
\caption{\textit{Left panel:} Ionizing rate (in per s) of EUV photons in the energy range $13.6\,\rm{eV} - \infty$ from the central protostar as a function of time for the RHD and RMHD runs. \textit{Right panel:} Heat released per ionization, $\mathcal{E}_{\nu}$, in the two runs. Solid curves are for ionization of H (energy range $13.6\,\rm{eV} - \infty$) whereas dashed curves denote ionization of H$_2$ (energy range $15.2\,\rm{eV} - \infty$). Lower accretion rates in the RMHD case lead to hotter and more compact stars, leading to a larger $\mathcal{E}_{\nu}$ as compared to the RHD case.}
\label{fig:ieuv_heat}
\end{figure*}

\subsection{Accretion discs and mass transport}
\label{s:results_accrates}
Having looked at the physical and chemical properties of the gas around the stars, we now turn to the accretion rates of the stars that form in the simulations and the physics behind it, which is crucial for interpreting the properties of the formed stars (\autoref{s:results_stellar}). We first look at the accretion rates of the isolated stars in the HD, RHD, and RMHD runs, and that of the oldest and most massive star in the MHD run. \autoref{fig:accr} plots the temporal evolution of the accretion rate averaged over $100\,\rm{yr}$ intervals for the different runs. All the accretion rates are quite high in the first $1000\,\rm{yr}$, and decline as time progresses; the accretion rate in the RMHD run approximately declines with time as $\dot M_{\rm{acc}} \propto t^{-0.6}$ for the duration of the simulation.

Interestingly, the long-lived accretion burst in the MHD run at $t \approx 1500\,\rm{yr}$ is followed by the onset of fragmentation, as we show via the dashed orange curve in \autoref{fig:accr}. There is also a clear difference between runs that include magnetic fields and those that do not: both the MHD and RMHD runs exhibit lower accretion rates than the HD and RHD runs by a factor of few \textit{at all times}. In fact, the most massive star in the MHD run undergoes multiple periods of quiescence due to fragmentation. As we shall see later in \autoref{s:results_stellar}, this has significant consequences for the masses of the stars that form in the presence of magnetic fields. 

To further understand why the accretion rates onto the star are lower in runs with magnetic fields, we look at the temporal evolution of the available mass budget over multiple scales that supply mass to the star. The top left panel of \autoref{fig:disks} plots the mass within the envelope that constitutes a spherical region of radius $0.01\,\rm{pc}$ centered on the star; this envelope feeds the accretion disc of the star via infalling gas. We see from these panels that the amount of mass reaching the envelope initially increases in the MHD and RMHD runs, but then turns over and starts to decline. In contrast, the rate of mass transfer from the envelope to the accretion disc is initially quite fast in the HD and RHD runs, resulting in a decline of mass in the envelope and buildup of mass in the disc, as we see from the top right panel of \autoref{fig:disks}. This mass is consequently accreted by the star at a high rate. The mass in the envelope of the HD and RHD runs is initially quite similar, but starts to diverge beyond $1000\,\rm{yr}$. As we will see in \autoref{s:results_accrates_mass}, this epoch coincides with a decline in the mass growth of the star in the RHD run as compared to the HD run. This occurs because radiation pressure slows down gas infall onto the star, which leaves more mass in the envelope. Thus, even though ionizing feedback is weak and acts close to the star, the non-linear connection between different scales leads to noticeable changes in the mass budget at scales much larger than the star. Since the mass of the star is initially small, the ratio $M_{\rm{disc}} / M_{\star}$ is high (cf. bottom left panel of \autoref{fig:disks}). We learn that magnetic fields limit the amount of gas infalling onto the envelope at later stages, leading to mass depletion within the accretion disc. However, in addition to this, the rate of mass transport from the envelope to the disc also declines in the presence of magnetic fields, as we show in the bottom right panel of \autoref{fig:disks}. There is no further fragmentation in the HD, RHD and RMHD runs because the mass in the disc continues to accrete efficiently onto the central star, thereby ensuring $M_{\rm{disc}} / M_{\star} < 1$. We demarcate the onset of fragmentation in the MHD run by dashed curves in all the panels. The mass transport from the envelope to the disc is so quick in the MHD run that the disc becomes $10\times$ as massive as the central star, and fragments. Subsequent peaks in $M_{\rm{disc}}/M_{\star}$ at late times mark the onset of further fragmentation within the accretion disc of the oldest star.

To summarise, the mass growth of the stars in our simulations are governed by the rate of mass transfer from larger scales onto the envelope, envelope to the accretion disc, and accretion disc to the star. At each stage of mass transfer, magnetic fields significantly influence the outcome and the total mass budget available. We also learn that, in addition to this transport, fragmentation is governed by whether the accretion disc becomes too massive as compared to the central star (\textit{i.e.,} $M_{\rm{disc}} / M_{\star} \gg 1$). This can occur if the accretion rate onto the star is small, for example, due to protostellar feedback or negative feedback from fragmentation itself.

\subsection{Stellar properties}
\label{s:results_stellar}

\subsubsection{Stellar mass}
\label{s:results_accrates_mass}
Next, we look at the evolution of stellar mass of the stars formed in the four runs in \autoref{fig:mass}. We plot two curves for the MHD run: the solid orange curve shows the mass accumulated in all the stars that form, whereas the dashed orange curve plots the mass growth for the oldest and the most massive star. The mass growth of the oldest and the most massive protostar in the MHD and RMHD runs closely track each other, and are substantially slower as compared to the HD and RHD runs. However, the total mass in stars in the MHD run is almost the same as that in the HD and RHD runs. In the HD and RHD runs, we find that the mass growth roughly follows $M_{\star} \propto t^{0.57}$ whereas in the MHD and RMHD runs, it follows $M_{\star} \propto t^{0.47}$. At the time we stop the simulations, the stellar masses in the four runs are: $127\,\rm{M_{\odot}}$ (HD), $48\,\rm{M_{\odot}}$ (MHD), $118\,\rm{M_{\odot}}$ (RHD), and $57\,\rm{M_{\odot}}$ (RMHD).

It is clear that magnetic fields have a dramatic impact on the stellar mass of Population III stars. Interestingly, the RMHD run is very distinct from the other runs in that the total mass in stars is a factor of two lower. Thus, the combination of magnetic fields and ionizing feedback acts to lower the star formation efficiency. It is also noteworthy that within the first $5000,\rm{yr}$ of star formation, we already observe a twofold difference in the stellar mass between the RHD and RMHD runs. We also find that magnetic fields play a much more important role than ionizing feedback at the early epochs ($\lesssim 5000 \, \mathrm{yr}$) of protostellar evolution that we simulate in this study. Ionizing feedback has a limited impact in lowering the stellar mass, even when we average accretion rates over some timescale to limit numerical noise in the stellar parameters (see \aref{s:app_averaging}). If radiation feedback becomes dominant at later times, it will suppress mass growth of the star, further limiting the maximum mass a Population III star can acquire. This finding has important implications for the IMF of Population III stars.

\subsubsection{Stellar radius and effective temperature}
The mass of a star and the accretion rates onto it affect the evolution of the properties of the protostar via the subgrid model we adopt (cf. \autoref{fig:ieuv}). This implies that the differences in these quantities (mass/accretion rates) we present in previous sections would have direct consequences on the stellar properties, and thereby their radiative outputs. To investigate this, we study the evolution of the protostellar radius, $R_{\star}$, and effective temperature, $T_{\rm{eff}}$, in these runs. \autoref{fig:radius_teff} presents the results. We find that the star in the RMHD run is more compact and bluer than that in the RHD run, which can be attributed to its lower accretion rate (\autoref{fig:accr}). However, some of the variation we see in the stellar properties can be a result of the noise in the accretion rate, which depends on the grid resolution. A common approach to mitigate this issue is to average the instantaneous accretion rates over some timescale to suppress numerical noise \citep[e.g.,][]{2016MNRAS.462.1307S}. We find that averaging the accretion rates this way to calculate stellar properties has a negligible impact (see \aref{s:app_averaging}), and our finding that the RMHD run produces more compact and bluer stars holds regardless.

$R_{\star}$ and $T_{\rm{eff}}$ set the overall ionizing output and heating rate due to ionization. To explore the impact of more compact and bluer stars as in the RMHD run, we plot the rate of ionizing photons released, $\mathcal{I}_{\rm{EUV}}$, and associated thermal energy deposition in the gas per ionization event, $\mathcal{E}_{\nu}$, in the left and right panels of \autoref{fig:ieuv_heat} respectively. 
We find that there is no considerable difference in $\mathcal{I}_{\rm{EUV}}$ after $2500\,\rm{yr}$, but $\mathcal{E}_{\nu}$ is higher in the RMHD run for the entire duration of the simulation. This is not surprising since $\mathcal{I}_{\rm{EUV}}$ depends on both $T_{\rm{eff}}$ and $R_{\star}$ whereas $\mathcal{E}_{\nu}$ depends only on $R_{\star}$: larger $R_{\star}$ but lower $T_{\rm{eff}}$ in the RHD runs seems to counterbalance the lower $R_{\star}$ but higher $T_{\rm{eff}}$ in the RMHD ones. The larger heating in the RMHD run is partially responsible for rendering the gas in the immediate vicinity of the star hotter, as we see in \autoref{fig:radialprofiles}.

\begin{figure}
\includegraphics[width=\columnwidth]{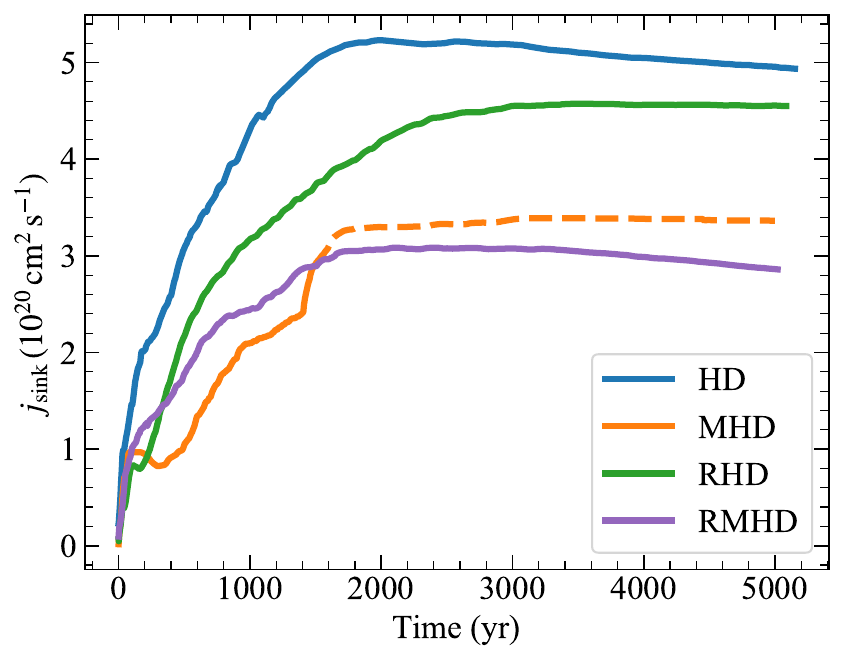}
\caption{Specific angular momentum of the sink particle in the different simulations, as a function of time. Dashed orange curve marks the onset of fragmentation in the MHD run. Magnetic fields redistribute angular momentum, and therefore lead to less angular momentum than in the case without magnetic fields. This can lead to Population III stars rotating at a smaller fraction of the breakup speed than anticipated.}
\label{fig:j}
\end{figure}

\subsubsection{Stellar rotation}
\label{s:results_accrates_rotation}
Another quantity of interest we can infer from our simulations is the angular momentum budget of the protostars, which is fundamental to understanding whether Population III stars were fast rotators. This is important for a number of observational probes -- most notably, strong rotation will induce mixing that can dredge up certain elements from the stellar interior to the surface \citep[][]{yoon2012,murphy2021,2023MNRAS.526.4467J,2024ApJS..270...28R,2024arXiv240416512T,2024arXiv240511235N}, which can enrich the surrounding ISM via stellar winds or supernovae \citep[e.g.,][]{2002ApJ...567..532H,2010ApJ...724..341H,2018ApJ...857...46I}. Rotation also plays a key role in determining whether Population III stars produced gamma ray bursts (GRBs) at the end of their lifetime \citep[e.g.,][]{2012ApJ...760...27W,2024ApJ...961...89V}. Almost all previous work on the rotation rate of Population III stars conclude that they were fast rotators, and rotated close to breakup speeds (see \citealt{hirano2018_rotation} for an exception).

Our simulations track the angular momentum budget of sink particles; however, we cannot directly interpret this as the angular momentum of the protostar, since the latter would be affected by gas dynamics at the sub-grid level. Some authors have attempted to quantify rotation by assuming the disc is in a Keplerian orbit and all the mass accreted by the sink will end up in a star \citep[e.g.,][]{stacy2011_rotation}, however this approach carries significant uncertainty due to the lack of information at the sub-au scale.
Instead of deriving the rotation speed of the star, we use our simulations to study the relative trends between the HD, RHD, MHD and RMHD runs, as is often done for Population I star formation \citep[e.g.,][]{2024A&A...690A.272K}.

\autoref{fig:j} plots the angular momentum of the most massive sink particle, $j_{\rm{sink}}$, as a function of time for the four simulations. We find that $j_{\rm{sink}}$ initially increases in all runs, as the angular momentum of the rapidly accreting material is added onto the sink. It eventually saturates (or starts mildly declining) as the accretion rates go down and the stellar mass goes up. It is also evident that $j_{\rm{sink}}$ is lower in the MHD and RMHD runs, owing to smaller central stellar masses and lower accretion rates.

Thus, the overall impact of magnetic fields, as expected from semi-analytical models \citep{hirano2018_rotation}, is to slow down the rotation of Population III stars, similar to their role in setting the rotation rates of Population I stars. If Population III stars could eject material via protostellar outflows \citep[][]{2006ApJ...647L...1M,2008ApJ...685..690M,2009ApJ...703.1096S,2013MNRAS.435.3283M}, the rotation rates would be even lower, as in Population I star formation \citep[e.g.,][]{2004A&A...423....1J,2021MNRAS.507.2448M}. However, the generation of strong large-scale mean field required for such outflows (e.g., via the large-scale $\alpha-\Omega$ dynamo -- \citealt{1981MNRAS.195..881P,1981MNRAS.195..897P,2005PhR...417....1B}) has not yet been conclusively shown in any existing Population III star formation simulations \citep[e.g.,][]{2019arXiv191107898L,2021MNRAS.503.2014S}.

\begin{figure}
\includegraphics[width=\columnwidth]{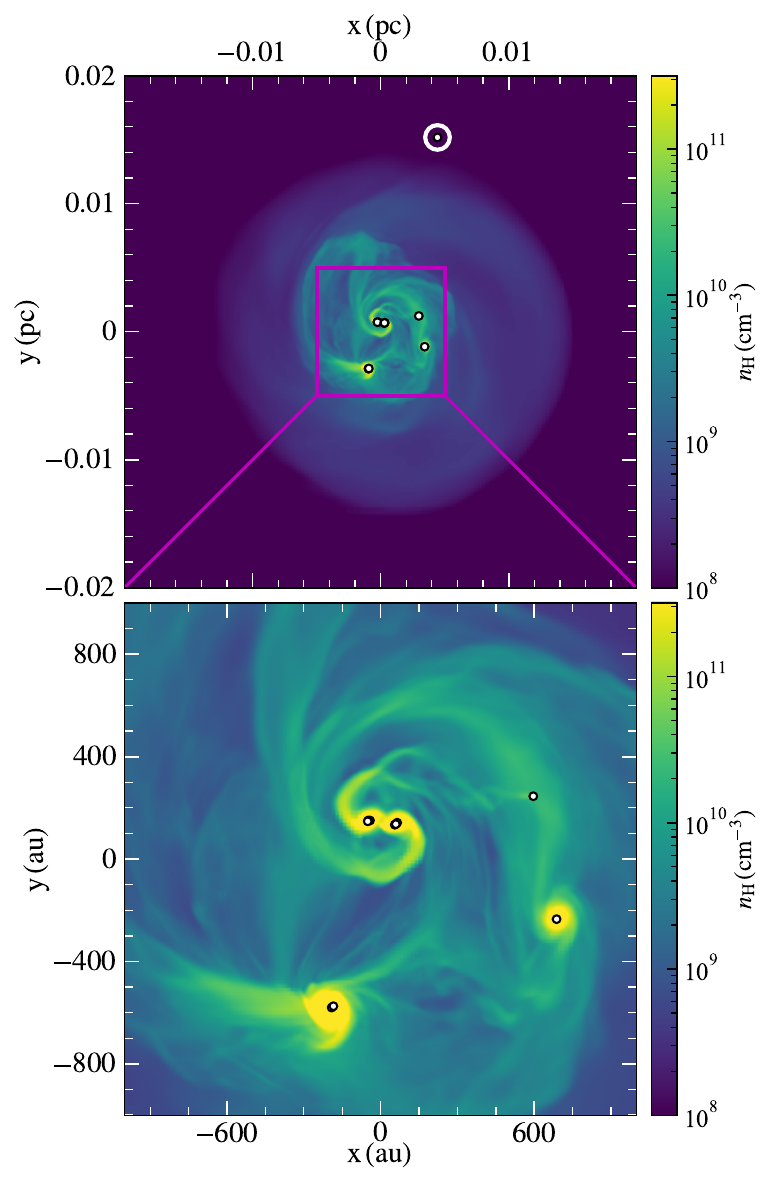}
\caption{\textit{Top panel:} density-weighted projection (along the $\hat{z}$ axis) of the gas density in one of the two turbulent RMHD realizations that fragment to produce Population III star clusters. White dots mark the position of all the stars. One of the (sub-Solar) mass stars is ejected from the cluster due to N-body interactions, as indicated by the white circle at its position. This star has a mass $0.45\,\rm{M_{\odot}}$, and stopped accreting after it was ejected. \textit{Bottom panel:} zoom-in of the central $0.01\,\rm{pc}$ region that harbours the cluster of 10 stars. Several stars are unresolved in this plot as they are located very close to other stars. Multiple stars have their respective accretion discs interconnected via dense filamentary structures.}
\label{fig:1seed_zoompanels}
\end{figure}

\begin{figure*}
\includegraphics[width=\textwidth]{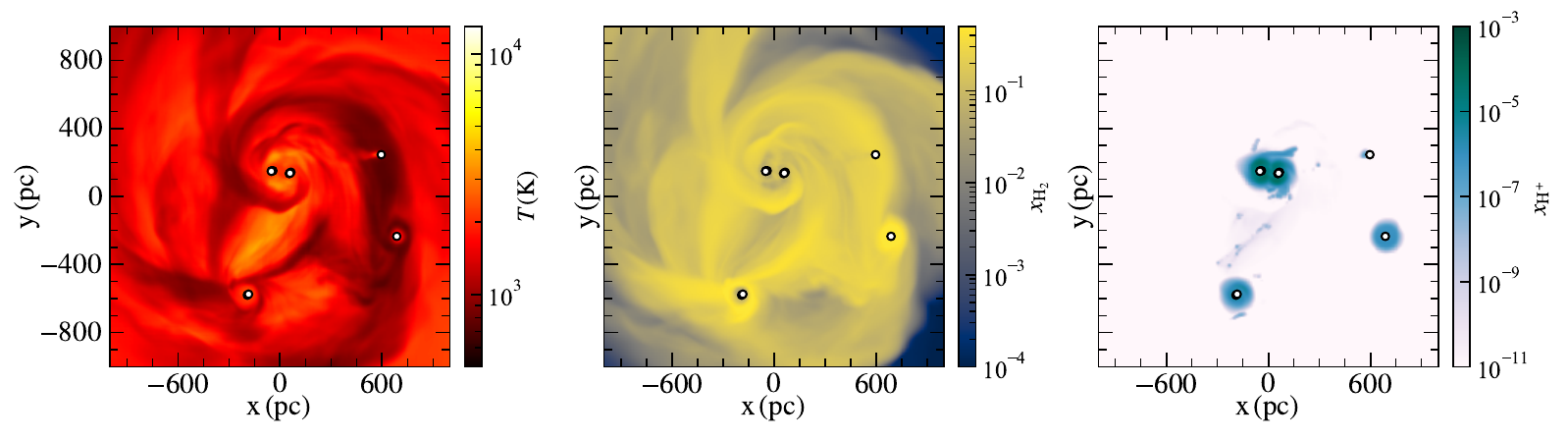}
\caption{Same as the bottom panel of \autoref{fig:1seed_zoompanels} but for the density-weighted gas temperature, and mass fractions of H$_2$ and H$^+$. Several stars in the cluster have partially ionized \ion{H}{ii} regions around them, due to lower accretion rates. A vast majority of the gas remains molecular due to otherwise colder temperatures.}
\label{fig:1seed_3panels}
\end{figure*}

\begin{figure}
\includegraphics[width=\columnwidth]{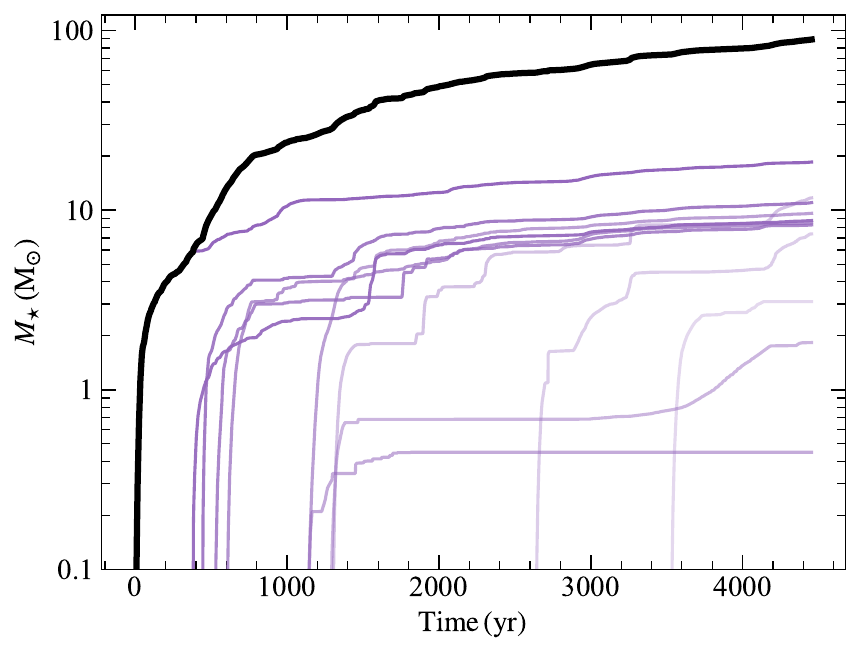}
\caption{Mass growth of all the 11 protostars in same RMHD run as in \autoref{fig:1seed_3panels}. Solid black curve depicts the total stellar mass of all stars. Progressively faded purple curves represent younger protostars, which can also be read off from the abscissa. The accretion rates for all stars are lower than the typical accretion rate seen in the isolated star run (cf. \autoref{fig:accr}). The $0.45\,\rm{M_{\odot}}$ star that gets ejected from the cluster by the end of simulation is born at $t \approx 1200\,\rm{yr}$. It stops accreting when it is ejected, at $t \approx 1700\,\rm{yr}$.}
\label{fig:1seed_mgrow}
\end{figure}

\begin{figure}
\includegraphics[width=\columnwidth]{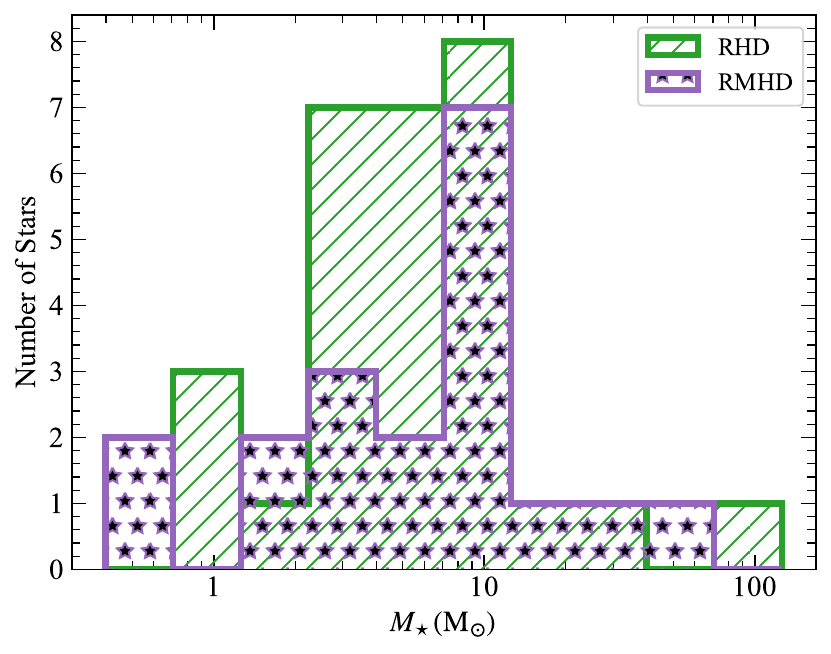}
\caption{Mass distribution of stars in the three RHD (green) and RMHD (purple) turbulent realizations at the end of the simulations.}
\label{fig:1seed_pdf}
\end{figure}

\section{Evolution of Population III clusters}
\label{s:results_cluster}
As we mention in \autoref{s:setup}, whether fragmentation will occur or not is sensitive to the turbulent realization used. Having looked at the simplest realization that (mostly) resulted in a single star, we now look at the evolution of stellar masses in the other RMHD realizations that produced star clusters. We only focus on the RMHD runs for this purpose, since it is the most physically realistic run.

\autoref{fig:1seed_zoompanels} and \autoref{fig:1seed_3panels} plot the density-weighted projections of the gas density, temperature and mass fraction of H$_2$, centered on the center of mass of all stars in one of the realizations. The zoom-in view of the central $0.01\,\rm{pc}$ of the cluster in \autoref{fig:1seed_zoompanels} shows how several stars are located very close to each other, but some stars are also located farther away in smaller, bound systems. We see the presence of cold, molecular gas throughout the central region, similar to that in the RMHD realization that produced an isolated star. The accretion discs of stars are interconnected via dense filamentary structures. This realization produces 11 stars within the first $5000\,\rm{yr}$ since fragmentation. The total stellar mass of all the stars is $\approx 90\,\rm{M_{\odot}}$ (solid black curve in \autoref{fig:1seed_mgrow}), however, the masses vary between $0.45 - 18.5\,\rm{M_{\odot}}$. The maximum mass is thus much lower than the total stellar mass, and also lower than the first RMHD realization because of fragmentation-induced starvation that lowers the accretion rate onto the stars, as we read off from \autoref{fig:1seed_mgrow}. This is a well known outcome of fragmentation \citep[e.g.,][]{2010ApJ...725..134P,Krumholz11e,2012MNRAS.420..613G,2022MNRAS.510.4019P,2022MNRAS.509.1959S}.

One implication of the low accretion rates is that the ionizing output from these stars is consistently higher as compared to the isolated stars we study above in \autoref{s:results}. To visualize the effects of ionizing feedback, we plot density-weighted projections of the mass fraction of H$^+$ in \autoref{fig:1seed_3panels}. The average mass fraction of H$^+$ only goes as high as $10^{-3}$, similar to that we see in the isolated star run. However, the size of the partially ionized \ion{H}{ii} region is larger. The gas temperature in these H$^+$ bubbles is also visibly higher, but remains below $10^{4}\,\rm{K}$. This increased ionizing feedback aids magnetic fields by further suppressing accretion onto these stars.

It is also interesting to identify if any of the stars that form in the cluster are part of binaries or higher order systems. We follow the algorithm developed by \citet{2009MNRAS.392.1363B} to calculate the multiplicity of stars formed in the cluster. This algorithm recursively finds pairs of most bound stars, and builds up stellar configurations upto quadruples until no more bound stars are left. We only consider bound configurations as high as quadruples since higher order stellar systems are very likely to disintegrate into smaller order systems. 

We find that the cluster (11 stars) we show in \autoref{fig:1seed_zoompanels} consists of one quadruple, one triple, and one binary system, whereas the remaining two stars are isolated. We cannot comment on the lifetime of such close multiples beyond the period we simulate, but note that the survival of such systems has several important implications for interactions, mass transfer, mergers and gravitational waves \citep[][]{2023MNRAS.524..307S,2024A&A...690A.106M}. The two stars that are isolated also happen to be the ones with the lowest masses, consistent with the general expectation that the multiplicity fraction increases with increasing stellar mass \citep[e.g.,][]{2012MNRAS.419.3115B,2012ApJ...754...71K,2020MNRAS.497..336S,2021MNRAS.507.2448M}. Of the two isolated stars, one gets ejected out of the cluster due to N-body interactions; we highlight this with the white circle in the projections in \autoref{fig:1seed_zoompanels}. This star is born at $t \approx 1200\,\rm{yr}$, and stops accreting once it leaves the cluster at $t \approx 1700\,\rm{yr}$. We can see in \autoref{fig:1seed_mgrow} that the mass of this star at the time it leaves the cluster is $0.45\,\rm{M_{\odot}}$, and it does not increase for the remainder of the simulation. It is very unlikely that the mass of this star will increase considerably after this period, even if accretion onto the minihalo continues, as the accreted gas will flow down the potential well to the center of the minihalo rather than to the (low-mass) star (see also, \citealt{2022MNRAS.510.4019P,2022ApJ...925...28L}). Such low mass Population III stars could have potentially survived to the present day \citep{2001A&A...371..152M}. However, they form rarely and only remain sub-Solar if they are ejected from the cluster, so the chances of observing these stars are dim. In the third turbulent RMHD realization, the star cluster consists of 7 stars, out of which two are locked in a binary and four in a quadruple configuration (see \aref{s:app_rmhd}). The mass of the oldest and the most massive star is $30\,\rm{M_{\odot}}$. The star with the lowest mass ($0.7\,\rm{M_{\odot}}$) remains isolated in this realization as well, but it continues to accrete by the time we stop the simulation as it does not get ejected.

Finally, we look at the distribution of stellar masses in all the three RHD and RMHD realizations (including the first realization which only forms one star). \autoref{fig:1seed_pdf} plots the results. The RHD runs form a total of 30 stars whereas the RMHD run forms 20 stars. The stellar masses span a wide range, between $0.45 - 117\,\rm{M_{\odot}}$. There is no discernible difference between the two mass distributions on average, which is not unexpected given the low number statistics. However, despite the limited numbers, our results suggest that forming $>100\,\rm{M_{\odot}}$ Population III stars is very challenging in the presence of strong magnetic fields, especially in cases where the cloud fragments to produce multiple stars. Recent simulations by \citet{2022MNRAS.510.4019P} that do not include all the physics but resolve gas collapse to much higher densities also reach similar conclusions on the maximum mass of Population III stars. One potential way around this challenge is for the minihalo in which the cloud is embedded to continuously accrete material from the cosmic web. However, the outcome in such a scenario is still unexplored since achieving sufficient resolution in a large volume simulation for long time periods have not yet proven computationally feasible.

\section{Caveats}
\label{s:caveats}

\subsection{Physical limitations}

\subsubsection{Dissociating feedback}
The key stellar feedback physics missing from our RMHD simulations is dissociating feedback due to H$_2$ dissociation by photons in the Far-UV Lyman-Werner band ($11.2 - 13.6\,\mathrm{eV}$). These photons dissociate H$_2$ into H, and can bring down the overall H$_2$ cooling rate, which can possibly make the gas hotter than that due to ionizations only. Only a handful of 3D simulations exist that include both ionizing and dissociating radiation feedback \citep{2016ApJ...824..119H,2020ApJ...892L..14S,2023ApJ...959...17S,2022MNRAS.512..116J}, out of which only \citet{2016ApJ...824..119H} and \citet{2023ApJ...959...17S} carry out a thorough comparison of the effects of ionizing and dissociating feedback. These authors find that FUV (dissociating) feedback can act on low density gas much farther out from the protostars, leading to H$_2$ dissociation in the envelope that in turn increases the gas temperature and decreases the rate of accretion onto existing stars. However, fully terminating accretion onto protostars requires EUV (ionizing) feedback from expanding \ion{H}{ii} regions. The overall mass growth is therefore much slower in their EUV+FUV runs as compared to the FUV only run. As \cite{2022MNRAS.512..116J} point out, however, these conclusions are subject to resolution and implementation of feedback (e.g., whether photons are injected at the location of the sink particle or at the edge of the accretion radius). Further, none of the simulations above include shielding of H$_2$ by H, which is expected to be important in high density regions where the column density of H can be very high \citep{2011MNRAS.418..838W,2023ARA&A..61...65K}. In a recent work, we have incorporated FUV feedback \citep{2025arXiv250112734S}, finding similar conclusions on the role of magnetic fields at early times.

\subsubsection{Lyman-$\alpha$ radiation pressure}
We also do not model radiation pressure due to resonant scattering of Lyman-$\alpha$ photons within the accretion disc \citep[e.g.,][]{2002ApJ...569..558O,2008ApJ...681..771M}, as it requires modeling line-driven radiative transfer which is beyond the scope of our numerical implementation of radiative transfer. Estimates suggest that the momentum imparted could be as large as a $1000 \times$ the available radiation pressure from ionizing photons, due to high optical depths which results in multiple scatterings \citep{2022MNRAS.512..116J}. This momentum can counteract the gravitational influence of the star and permit \ion{H}{II} regions to break out of the disc by reducing the density in its immediate vicinity. However, subtleties such as velocity gradients and collision-induced Lyman-$\alpha$ suppression due to 2p$\rightarrow$2s transitions can suppress the force multiplication factor significantly \citep{2025MNRAS.tmp...39N}, so the role of Lyman-$\alpha$ radiation pressure in Population III star remains to be explored.

\subsubsection{Non-ideal MHD}
Our MHD and RMHD simulations do not include non-ideal MHD effects. Non-ideal MHD effects are critical in forming accretion discs around Population I stars and regulating mass and angular momentum transport \citep[see the review by][]{2018FrASS...5...39W}. Out of the three manifestations of non-ideal effects (Ohmic dissipation, ambipolar diffusion, Hall effect), diffusivity due to ambipolar diffusion during Population III star formation has been shown to be orders of magnitude higher than the other two \citep{2012ApJ...754...99S,2019MNRAS.488.1846N,2020MNRAS.496.5528M,2023MNRAS.519.3076S}. The general result of these studies is that ambipolar diffusion can slightly suppress but cannot halt the growth of magnetic fields via dynamo amplification, and strong magnetic fields should be expected during Population III star formation \citep[see also, the discussion in ][section 3.4, which also applies to our work]{2021MNRAS.503.2014S}. However, the impact of non-ideal MHD effects close to the protostars, where ionizing feedback can yield a higher ionization fraction has not yet been investigated. Current primordial chemical networks typically exclude Li as it is not an important coolant \citep[e.g.,][]{1998A&A...335..403G,2013ARA&A..51..163G,2018MNRAS.476.1826L}, but Li could be potentially important for non-ideal MHD effects since it is the dominant charge carrier in primordial conditions. These physical considerations therefore warrant further investigation.

\subsubsection{Rotation and turbulence}
We include solid-body rotation and trans-sonic turbulence in our initial conditions, inspired from cosmological simulations. However, we do not experiment with variations in these parameters, and thus cannot comment on how initial rotation and turbulence impact Population III star formation. A handful of studies exist that have scanned through the expected parameter space for rotational and turbulent energies \citep{2011ApJ...727..110C,2018MNRAS.479..667R,2023MNRAS.518.4895R,2020MNRAS.494.1871W}, finding that increasing turbulence and decreasing rotational energy enhance fragmentation. However, these studies do not achieve the Jeans resolution required to correctly capture the chemical state of H$_2$ in shocked regions. The interplay between rotation, turbulence, magnetic fields and radiation feedback in Population III conditions therefore remains undiscovered. In future work, we plan to address this gap by sweeping across a range of initial rotational energies and turbulent Mach numbers.

\subsection{Numerical limitations}
There are two categories of resolution requirements for simulating Population III star formation in the presence of both magnetic fields and radiation feedback: the absolute maximum grid resolution and the resolution needed per Jeans length \textit{at fixed absolute resolution}. These are distinct numerical considerations, with the former determining the smallest physical scales one can resolve, and the latter affecting how many regions exist at these higher resolutions. As we describe in this work, the latter is crucial to resolve shock thermodynamics and amplification of magnetic fields via small-scale dynamo \citep[e.g.,][]{2011ApJ...731...62F,2012ApJ...745..154T,2021MNRAS.503.2014S,2022ApJ...935L..16H}, whereas the former is important to resolve the inner scale height of accretion discs around Population III stars to better resolve the expansion of \ion{H}{ii} regions \citep{2022MNRAS.512..116J} and amplification of magnetic fields via large-scale dynamos \citep{2021MNRAS.503.2014S}.

While we are able to sufficiently satisfy the latter resolution requirement by using 64 cells per Jeans length, we do not achieve the sub-au resolution needed to sufficiently resolve the inner accretion disc. On the other hand, simulations that achieve sub-au resolution \citep{2012MNRAS.424..399G,2022MNRAS.516.2223P,2024A&A...685A..31P,2022ApJ...935L..16H,2022MNRAS.512..116J} do not include both magnetic fields and radiation feedback, do not sufficiently resolve the Jeans length to properly capture the effects of shocks on primordial chemistry, and cannot follow the simulation for a long time post star formation. Therefore, it remains to be explored how magnetic fields and radiation feedback interact on small scales very close to the protostar, and whether this interaction matters for the large scale evolution and final masses of Population III stars.

\section{Summary}
\label{s:conclusions}
In this work, we present the first suite of radiation-magnetohydrodynamics (RMHD) simulations of Population III star formation which simultaneously includes both magnetic fields and ionizing radiation feedback. We use a novel implementation of chemistry-coupled RMHD in the grid-based AMR code FLASH \citep{2022MNRAS.512..401M,MS24}. We start from a $1\,\rm{pc}$ primordial cloud (embedded in a dark matter minihalo at $z = 30$) and resolve the structure of the accretion disc around protostars down to few $\rm{au}$. Importantly, we resolve the Jeans length by 64 cells throughout the simulation to accurately capture shocks and shock-driven chemistry.

To explore the interplay between magnetic fields and ionizing feedback, we explore four scenarios: a control, hydrodynamic simulation without magnetic fields or ionizing feedback (HD), including magnetic fields (MHD), or ionizing feedback (RHD), and one that includes both (RMHD). In line with previous works that show dynamically strong magnetic fields are expected during Population III star formation due to dynamos \citep{2012ApJ...754...99S,2012ApJ...745..154T,2021MNRAS.503.2014S,2022MNRAS.516.3130S,2022ApJ...935L..16H,2024A&A...684A.195D}, we include a saturated turbulent magnetic field in our initial conditions for the MHD and RMHD runs. We carry out three different turbulent realizations that start with identical initial conditions. In one of these realizations, we find that only a single Population III star is produced in the HD, RHD, and RMHD runs whereas the MHD run fragments to produce four stars. We stop the simulations at $5000\,\rm{yr}$ since the formation of the first star.

We summarize our key findings as follows:
\begin{itemize}
\item{Resolved shocks significantly impact primordial chemistry. When shocks are well resolved (as is the case when we use 64 cells per Jeans length), gas in the post-shock region remains hot because H$_2$ dissociates faster than the gas can cool. This effect can significantly deplete H$_2$ around Population III protostars, in line with previous work \citep{2021MNRAS.503.2014S}.}

\item Mass transfer from larger scales to the star is systematically slower in the presence of magnetic fields, which limits the maximum mass of Population III stars. Additionally, ionizing feedback from the protostar is weak because of 1.) high accretion rates in the earliest stages, and 2.) ionizing flux getting trapped in dense regions close to the protostar, in line with the findings of \citet{2022MNRAS.512..116J}. The ionizing flux from the star is only able to create a partially ionized \ion{H}{ii} region around the star, especially in the absence of fragmentation.

\item Accretion discs are colder and significantly more molecular in the presence of magnetic fields, likely because magnetic fields slow down gas compression and reduce compressional heating. This effect has also been observed in previous simulations \citep[e.g.,][]{2020MNRAS.497..336S,2022MNRAS.516.3130S,2022MNRAS.511.5042S}.

\item{Strong magnetic fields also redistribute angular momentum, producing sink particles with less spin. Consequently, this phenomenon is likely to result in Population III stars rotating at a somewhat slower rate than previously anticipated, in line with expectations from semi-analytical models \citep{hirano2018_rotation}.}

\end{itemize}

We thus conclude that magnetic fields are more important than ionizing feedback in regulating the mass growth of Population III stars during the earliest stages of Population III star formation. The mass of the star $5000\,\rm{yr}$ after the formation of the first protostar in the MHD and RMHD runs is only $\sim 50-60\,\rm{M_{\odot}}$, while HD and RHD runs produce stars with masses $> 100\,\rm{M_{\odot}}$. In the two other turbulent realizations, we observe widespread fragmentation in the RMHD runs, so there is clearly scope for doing several simulations to achieve statistically significant results and construct an IMF. Nevertheless, our general result that magnetic fields limit the maximum mass of Population III stars stands valid, since fragmentation further limits the stellar mass due to competitive accretion: in these runs, the maximum stellar mass within the same time period is only $\sim 20 - 30\,\rm{M_{\odot}}$. 

It remains to be seen if these differences at early times have a profound impact on the final mass and fate of Pop~III stars. Since magnetic fields limit the mass growth of the most massive stars at early times, they can induce stronger radiation feedback earlier than expected as the accretion rates decrease. Stronger feedback can further act to halt accretion and limit the maximum mass, thereby creating less massive Pop~III stars. Thus, the role of magnetic fields should not be ignored in Population III star formation.

\section*{Acknowledgements}
We thank Roman Gerasimov, Michael Grudić, Rachel Somerville, Blakesley Burkhart, Vanesa Diaz, Joakim Rosdahl, and Volker Bromm for useful discussions. We are grateful to the anonymous reviewer for their feedback that significantly improved this work. We thank Lionel Haemmerlé for sharing their Population III protostellar evolution model grids. PS is supported by the Leiden University Oort Fellowship and the International Astronomical Union -- Gruber Foundation (TGF) Fellowship. SHM acknowledges support through NASA grant No. 80NSSC20K0500 and NSF grant AST-2009679, and the Centre for Computational Astrophysics (CCA) at the Flatiron Institute. The simulations and data analyses presented in this work used high-performance computing resources provided by CCA, as well as by the Australian National Computational Infrastructure (NCI) through project \texttt{jh2} in the framework of the National Computational Merit Allocation Scheme and the Australian National University (ANU) Allocation Scheme, and through project \texttt{iv23} as part of a contribution by NCI to the ARC Centre of Excellence for All Sky Astrophysics in 3 Dimensions (ASTRO 3D, CE170100013). This work was also supported by the Dutch National Supercomputing Facility SURF via project grant EINF-8292 on \texttt{Snellius}. This work was performed in part at Aspen Center for Physics, which is supported by National Science Foundation grant PHY-2210452. We acknowledge using the following softwares: FLASH \citep{2000ApJS..131..273F,Dubey_2013}, \texttt{VETTAM} \citep{2022MNRAS.512..401M}, \texttt{KROME} \citep{2014MNRAS.439.2386G}, \texttt{petsc} \citep{petsc-efficient,petscsf2022}, Astropy \citep{2013A&A...558A..33A,2018AJ....156..123A,2022ApJ...935..167A}, Numpy \citep{oliphant2006guide,2020arXiv200610256H}, Scipy \citep{2020NatMe..17..261V}, Matplotlib \citep{Hunter:2007}, \texttt{yt} \citep{2011ApJS..192....9T}, and \texttt{cmasher} \citep{2020JOSS....5.2004V}. This research has made extensive use of NASA's Astrophysics Data System Bibliographic Services (ADS).

\section*{Data Availability}
Movies associated with this article are available as supplementary (online only) material. To access the suite of simulations or properties of the stars, please contact the authors.



\bibliographystyle{mnras}
\bibliography{references} 


\appendix

\section{Effects of radiation pressure}
\label{s:app_radpres}

\begin{figure}
\includegraphics[width=\columnwidth]{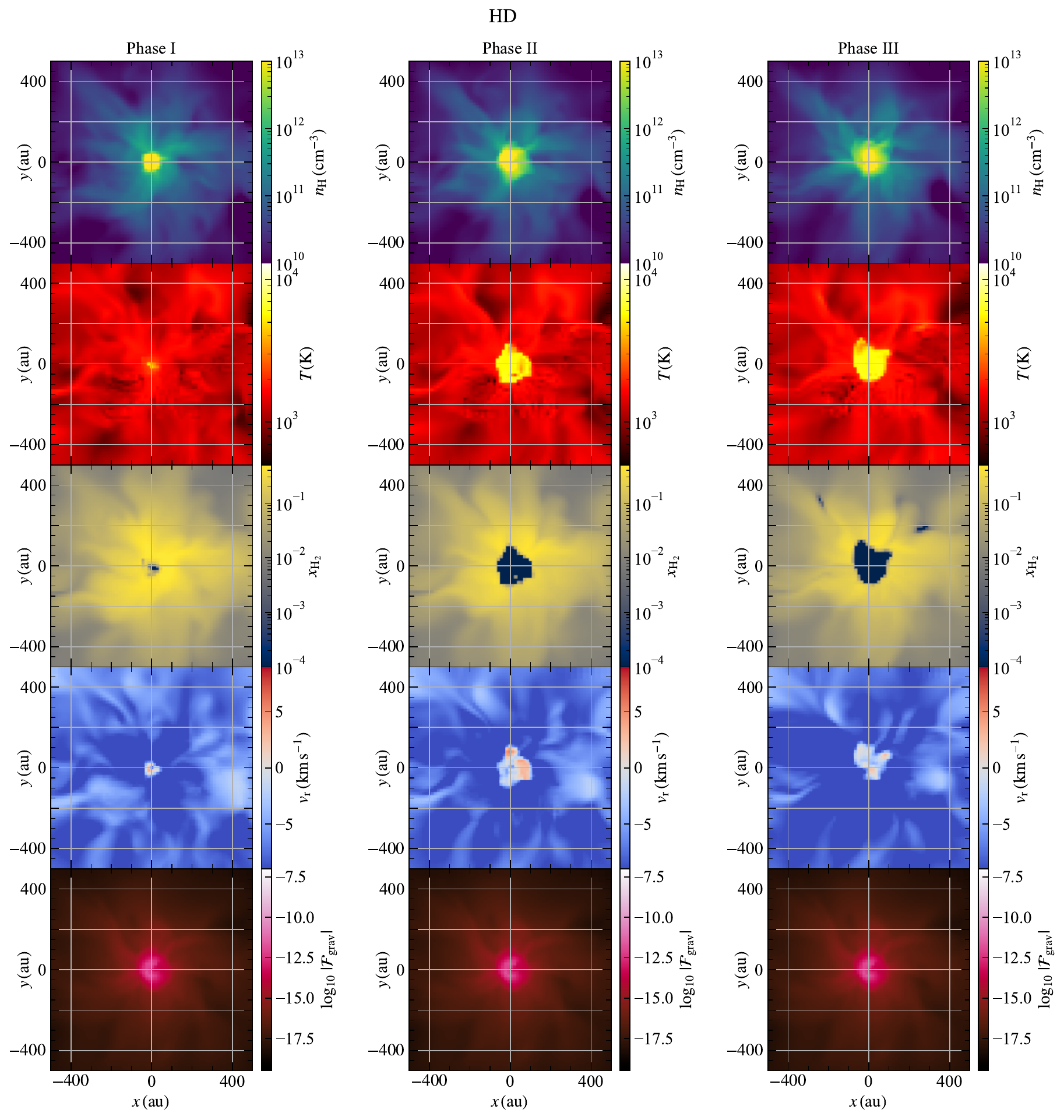}
\caption{\textit{Left panel:} 2D slices through the $\hat{z}$ axis of the gas density, temperature, radial velocity, and radial gravitational force per unit volume ($\mathcal{F}_{\rm{grav}}$) in the HD simulation just after star formation (termed Phase I). The slices are centered on the sink particle. \textit{Middle and right panels:} same slices as the left panel separated by $\sim 40\,\rm{yr}$ in the post shock regime, marked Phase II and III. Rapidly infalling gas (as depicted by large negative radial velocity) gets shocked as it approaches the sink particle, leading to shock heating, H$_2$ dissociation, and high gas temperatures in the central region.}
\label{fig:app_hdradp}
\end{figure}

\begin{figure}
\includegraphics[width=\columnwidth]{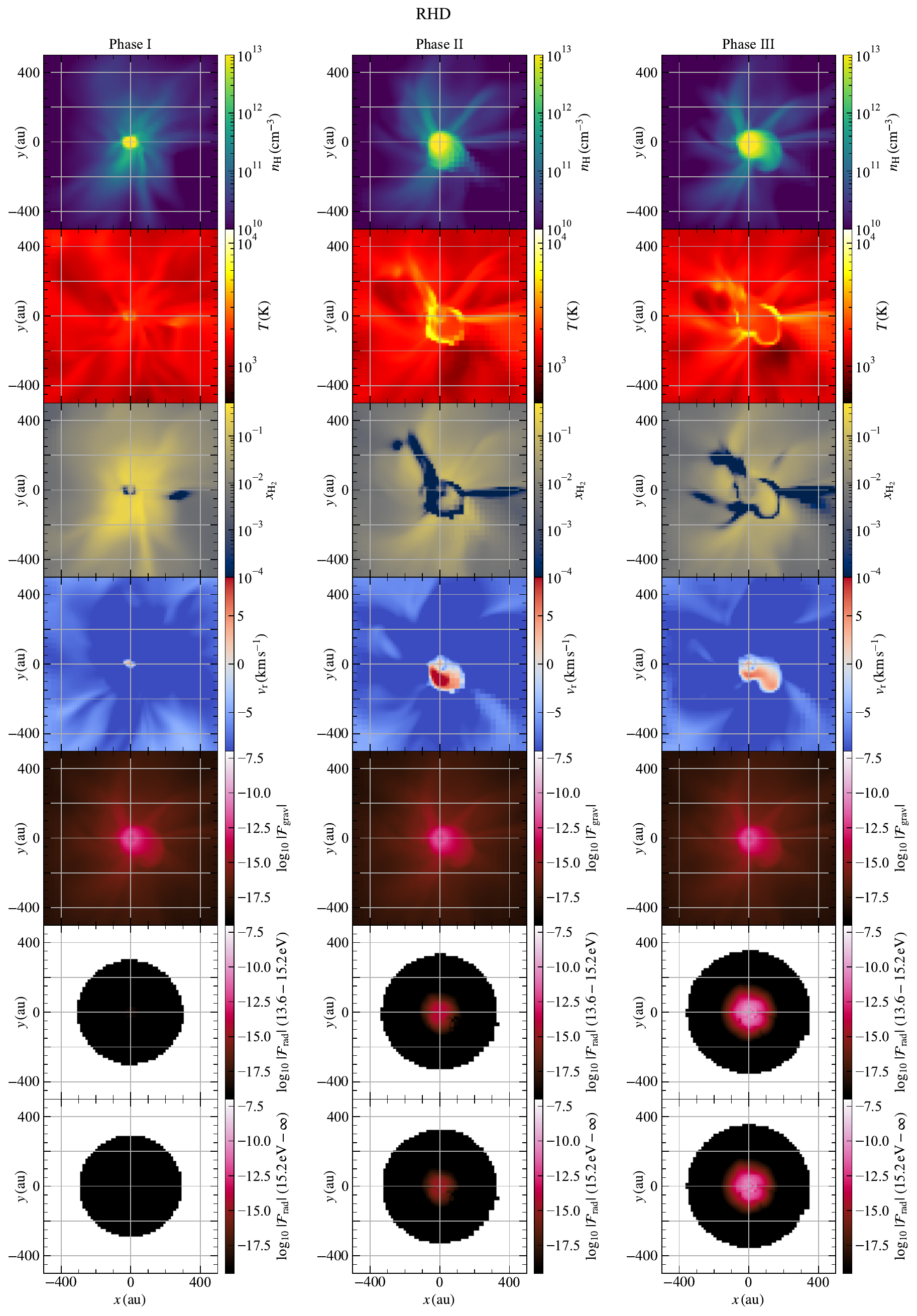}
\caption{\textit{Left panel:} 2D slices through the $\hat{z}$ axis of the gas density, temperature, radial velocity, radial gravitational force per unit volume ($\mathcal{F}_{\rm{grav}}$), as well as radial radiative force per unit volume ($\mathcal{F}_{\rm{rad}}$) for the H and H$_2$ ionizing bands in the RHD simulation. The slices are centered on the sink particle, and correspond to a time soon after the sink forms (Phase I). \textit{Middle and right panels:} same slices as the left panel separated by $\sim 40\,\rm{yr}$, called Phases II and III. Contrary to the HD simulation, radiation pressure counteracts gravitational force in the vicinity of the sink, pushing the gas outwards, which leads to large positive radial velocities. This effect in turn slows down gas infall and reduces shock heating, creating a low temperature zone surrounding the sink.}
\label{fig:app_rhdradp}
\end{figure}

Although the HD and RHD simulations yield similar stellar masses, the sink particle in the RHD simulation is surrounded by low temperature gas not seen in the HD simulation. We attribute this to radiation pressure imparted by ionizing photons, which acts against gravity and slows down gas compression near the sink. To quantify the effects of radiation pressure, we calculate the radial component of the radiative force per unit volume experienced by the gas as \citep[e.g.,][appendix C]{2008ApJ...681..771M}
\begin{equation}
\mathcal{F}_{\rm{rad}} = \frac{1}{c} \rho \kappa k_{\rm{H}} \left(\mathbf{F}\cdot \hat{r}\right)
\label{eq:s1}
\end{equation}
where $\rho$ is the gas volume density, $\kappa$ is the opacity of hydrogen atoms to the UV energy band photons considered (depends on the stellar effective temperature; for example, see \citealt{baczynski2015}), $k_{\rm{H}}$ is the number fraction of H, $\mathbf{F}$ is the radiation flux, and $\hat{r}$ the radial vector in the frame of reference of the star. We calculate $\mathcal{F}_{\rm{rad}}$ for both the H and H$_2$ ionizing energy bands (\textit{i.e.,} $13.6 - 15.2\,\rm{eV}$ and $15.2\,\rm{eV} - \infty $). The self gravity of the gas is negligible near the sink. So, we can express the radial component of the gravitational force per unit volume experienced by the gas as
\begin{equation}
\mathcal{F}_{\rm{grav}} = \rho \left(\mathbf{g}\cdot \hat{r}\right)
\label{eq:s2}
\end{equation}
where $\mathbf{g}$ is gas acceleration due to the gravity of the sink particle.

\autoref{fig:app_hdradp} plots key properties of the gas in the HD simulation at three different epochs (marked Phase I, II and III) separated by $\sim 40\,\rm{yr}$ soon after sink particle formation. The various panels show that gas rapidly falls onto the sink particle, leading to large negative radial velocities. This infall leads to shocks near the sink, and shock heating dissociates H$_2$ -- a phenomena explained in \citet[appendix A]{2021MNRAS.503.2014S}, which we highlight in \autoref{s:results_physical}. Since H$_2$ is the dominant gas coolant, removing it leads to high gas temperatures in subsequent phases, as we read off from the middle and right panels of \autoref{fig:app_hdradp}.

Let us now look at what happens in the RHD simulation using \autoref{fig:app_rhdradp}. Phase I in the RHD run resembles that in the HD run; the star has just appeared and does not yield significant radiation pressure. However, as the star grows its mass and releases ionizing photons, they exert momentum on the gas. The magnitude of $\mathcal{F}_{\rm{rad}}$ in both the ionizing bands becomes comparable to $\mathcal{F}_{\rm{grad}}$, so radiation pressure is able to counteract the influence of gravity and push the gas away from the sink. The Gaussian stellar injection term discussed in Section~\ref{s:setup_vettam} does smoothen the spatial zone over which radiation pressure is applied, but we expect similar outcomes to emerge as the total radial momentum injected (and the associated competition with gravity) is conserved. This is evident from the positive radial velocities in Phases II and III in the RHD simulation. Thus, weaker gas infall leads to a reduction in compressional heating, which ensures that the gas temperatures remain relatively lower and H$_2$ is not dissociated. We also see that the radius at which the gas gets shocked is located outside the zone of influence of radiation pressure, as expected. In this way, radiation pressure leads to qualitatively different outcomes for the physical properties of surrounding gas. However, as we discuss in the main text, this does not seem to considerably impact the mass growth of the star.

\section{Saturation of the small-scale dynamo}
\label{s:app_dynamo}
As we mention in \autoref{s:setup}, we include a saturated magnetic field in our initial conditions, inspired from previous works that find very fast, exponential amplification of weak field to near saturation levels due to the small-scale dynamo \citep{2010ApJ...721L.134S,2010A&A...522A.115S,2012ApJ...745..154T,2012ApJ...754...99S,2020MNRAS.497..336S,2021MNRAS.503.2014S,2023MNRAS.519.3076S,2024PASJ..tmp...61S,2024ApJ...962..158H}. However, this by itself does not guarantee that the field amplification we see later on is not due to a dynamo. This is an important check because if the field gets amplified by the small-scale dynamo later on, it will also play a role in shaping the stellar mass and accretion rate, which will lead to different conclusions about how magnetic fields influence Population III star formation.

\begin{figure}
\includegraphics[width=\columnwidth]{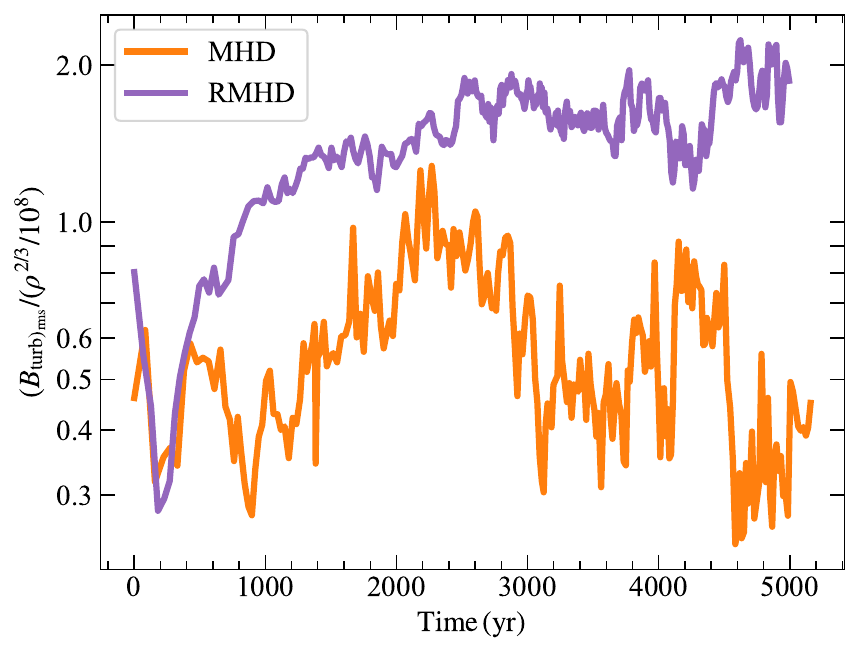}
\caption{Quantification of the small-scale dynamo action as a function of time in the MHD and RMHD runs. The ratio on the $y$ axis denotes the rms value of the turbulent magnetic field over density to the power two-thirds (to remove the contribution of flux-freezing to field growth), and normalized by $10^8$. The absence of an exponential growth of the small-scale dynamo indicates that the dynamo is saturated.}
\label{fig:app_dynamo}
\end{figure}

To confirm that the field remains saturated and the small-scale dynamo does not act, we calculate the strength of the turbulent component of the magnetic field, following the methodology laid out in \citet{2021MNRAS.503.2014S}. We then normalize the rms of the turbulent field, $(B_{\rm{turb}})_{\rm{rms}}$, to the gas density in order to remove contribution to field amplification from flux freezing. \autoref{fig:app_dynamo} shows the resulting time evolution of this quantity. We find that this quantity only changes within a factor of few during the course of the simulation, and does not exhibit an exponential rise as expected for the small-scale dynamo \citep[e.g.,][]{2011PhRvL.107k4504F,2012PhRvE..85b6303S,2015PhRvE..92b3010S}. Thus, the small-scale dynamo does not operate during the course of the simulation in both the MHD and RMHD runs, and the field remains close to saturation.

\section{Impact of averaging accretion rates}
\label{s:app_averaging}
In the main text, we used the instantaneous accretion rate onto the sink particle to calculate the stellar radius and temperature by interpolating across the stellar structure model grids from \citet{2018MNRAS.474.2757H}. However, this could lead to artefacts or unphysical variations in the stellar properties since the instantaneous accretion rate is sensitive to the numerical grid and resolution used. In the context of our work, instantaneous refers to accretion rates per timestep, which can be as low as 0.3 months.

To understand the impact this could have on our conclusions, we re-run the RHD simulation where we average the instantaneous accretion rate over some time period, and use the average rate while calculating the stellar properties. This approach has been demonstrated in previous Population III simulations \citep[e.g.,][]{2016MNRAS.462.1307S}. The time period over which the averaging can be carried out is somewhat ambiguous. Given that our average time resolution is of the order of $0.1\,\rm{yr}$, we average the accretion rates over $10\,\rm{yr}$ so that we can sufficiently smoothen out any small-term numerical variability while ensuring realistic episodic variations on timescales larger than $10\,\rm{yr}$ are adequately captured.

\begin{figure}
\includegraphics[width=\columnwidth]{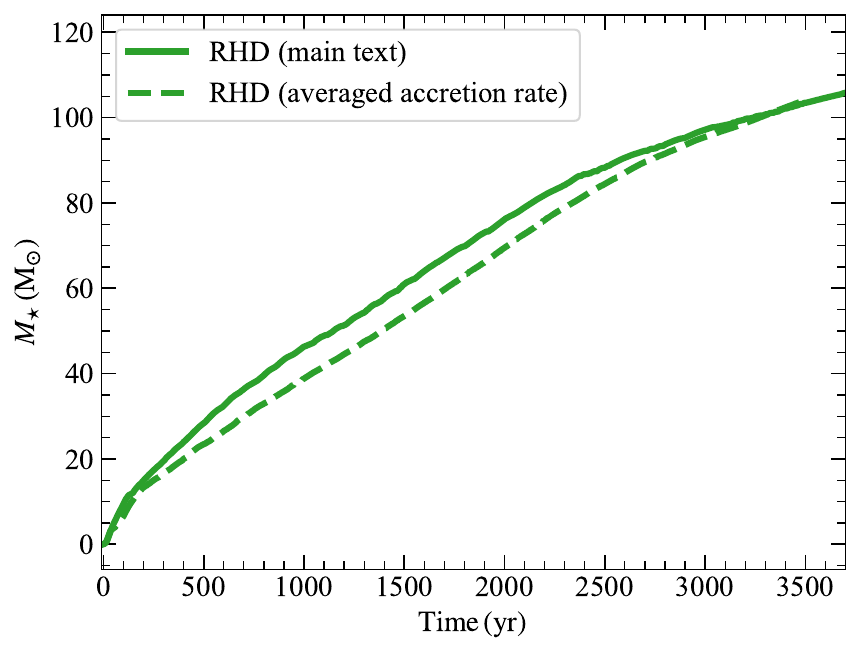}
\caption{Mass of the central star as a function of time in the RHD case presented in the main text (solid orange curve), compared with the alternate case where instantaneous accretion rates are averaged over a $10\,\rm{yr}$ time interval to reduce numerical noise in the rate of ionizing photons.}
\label{fig:app_averaging}
\end{figure}

\autoref{fig:app_averaging} plots the resulting protostellar mass as a function of time for this run as compared to that in the main text. The mass growth of the central protostar in the two simulations is similar, although the mass in the simulation using averaged accretion rates is lower by 10 per cent for a large duration. However, the ultimate mass in both cases is identical by $t = 3500\,\rm{yr}$. Therefore, we conclude that averaging the accretion rates has no impact on our results.

\section{Plots for the third RMHD turbulent realization}
\label{s:app_rmhd}
\begin{figure}
\includegraphics[width=0.7\columnwidth]{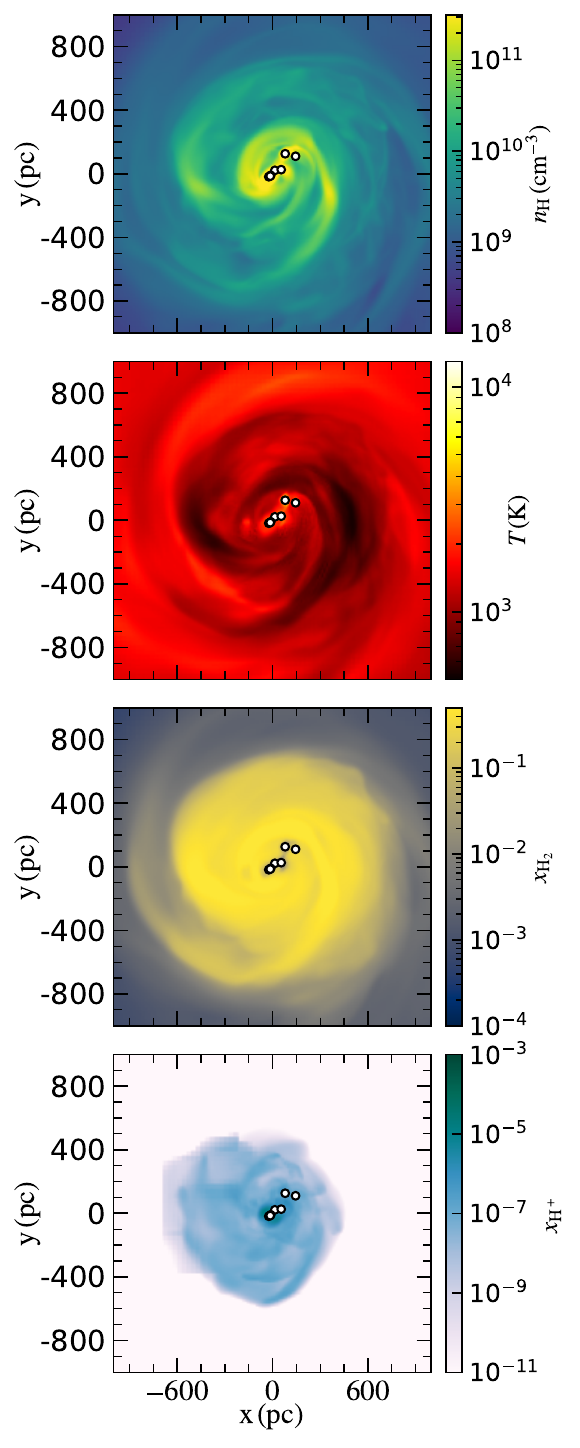}
\caption{Density-weighted projections of the gas density, temperature, and H$_2$ and H$^+$ mass fraction for the third RMHD turbulent realization. This realization forms 7 stars, with masses between $0.7 - 30\,\rm{M_{\odot}}$. The cluster remains intact in this realization as opposed to the second realization, leading to overlapping partially ionized \ion{H}{ii} regions.}
\label{fig:4seed_panels}
\end{figure}

\autoref{fig:4seed_panels} plots the density-weighted projections of the gas density, temperature, and mass fractions of H$_2$ and H$^+$ for the third RMHD turbulent realization. This realization creates 7 stars by the end of the simulation, none of which are ejected from the cluster. The minimum and maximum mass of stars in the cluster is $0.7$ and $30\,\rm{M_{\odot}}$. Similar to the other two turbulent RMHD realizations, this realization also exhibits cold, molecular gas due to strong magnetic fields. However, as opposed to the second realization, this cluster remains intact and the stars are close enough that their partially ionized \ion{H}{ii} regions overlap.

\label{lastpage}
\end{document}